\documentclass[useAMS,usenatbib,usegraphicx]{mn2e}

\usepackage{fixltx2e}

\def\apj{ApJ}

\def\aap{A\& A}

\def\araa{ARA\&A}
\def\mnras{MNRAS}

\def\mnras{{MNRAS}}

\def\pasp{{PASP}}

\def\Stotmath{{S_{\rm tot}}}

\def\'#1{\ifx#1i{\accent"13\i}\else{\accent"13#1}\fi}
\def\alamenos#1{$^{-#1}$}
\def\ala#1{$^{#1}$}

\def\diezala#1{10$^{#1}$}
\def\kms{km sec\alamenos 1}

\def\Npdf{$N$-pdf}
\def\Msun{$M_\odot$}

\def\SigmaA{$\Sigma-A_0$}

\def\be{\begin{equation}}
\def\ee{\end{equation}}

 \title[On the structure of molecular clouds] {On the structure of molecular clouds}

 \author[Ballesteros-Paredes, et al.]
 { \parbox{7.0in}{Javier Ballesteros-Paredes\ala 1 \thanks{e-mail:
        {\tt j.ballesteros@crya.unam.mx}}, Paola D'Alessio\ala 1 and Lee Hartmann\ala 2
} \\
\\
 \ala 1 Centro de Radioastronom\'ia y Astrof\'isica,
            Universidad Nacional Aut\'onoma de M\'exico, \\
            Apdo. Postal 72-3 (Xangari), Morelia,
            Michoc\'an 58089, M\'exico \\
\\
     \ala 2 Department of Astronomy, University of Michigan,  500
           Church Street, Ann Arbor, MI 48105, USA \\
}

\begin{document}

\date{Submitted to MNRAS, \today}

\pagerange{\pageref{firstpage}--\pageref{lastpage}} \pubyear{2012}
 
\maketitle

\label{firstpage}

\begin{abstract}

We show that the inter-cloud Larson scaling relation between mean
volume density and size $\rho\propto R^{-1}$, which in turn implies
that mass $M\propto R^2$, or that the column density $N$ is constant,
is an artifact of the observational methods used.  Specifically,
setting the column density threshold near or above the peak of the
column density probability distribution function \Npdf\ ($N\sim
10^{21}$~cm\alamenos 2) produces the Larson scaling as long as the
\Npdf\ decreases rapidly at higher column densities.  We argue that
the physical reasons behind local clouds to have this behavior are
that (1) this peak column density is near the value required to shield
CO from photodissociation in the solar neighborhood, and (2) gas at
higher column densities is rare because it is susceptible to
gravitational collapse into much smaller structures in specific small
regions of the cloud.  Similarly, we also use previous results to show
that if instead a threshold is set for the volume density, the density
will appear to be constant, implying thus that $M \propto R^3$.  Thus,
the Larson scaling relation does not provide much information on the
structure of molecular clouds, and does not imply either that clouds
are in Virial equilibrium, or have a universal structure.  {We
  also show that the slope of the $M-R$ curve for a single cloud,
  which transitions from near-to-flat values for large radii to
  $\alpha=2$ as a limiting case for small radii, depends on the
  properties of the \Npdf. }

\end{abstract}



\section{Introduction}\label{sec:intro}

More than 30 years ago, \citet{Larson81} published his
scaling relations for molecular clouds (MCs): the column
density-size relation (or ``Larson's third relation'')

\begin{equation}
  \rho \propto R^{\gamma_1}
\label{eq:larson_nr}
\end{equation}
and the velocity dispersion-size relation

\begin{equation}
  dv \propto R^{\gamma_2}.
\label{eq:larson_dvr}
\end{equation}
The exponents reported in that work were $\gamma_1=-1.1$ and
$\gamma_2=0.39$, respectively. However, the more widely accepted
values are $\gamma_1=-1$ and $\gamma_2=0.5$; and these have been
usually thought to be observational evidence that clouds are in Virial
equilibrium, although it is important to recall that any pair
$(\gamma_1,\gamma_2)$ satisfying $\gamma_1 = 2 \gamma_2 - 2 $ will be
consistent with Virial equilibrium \citep[see e.g.,][]{VSG95}.

Larson's relations have been used in many papers to describe the
internal structure of clouds \citep[e.g., ][ to cite just one of the
  more recent examples]{Goldbaum+11}.  However, their validity has
been called into question for observational reasons \citep{Kegel89,
  Scalo90, VBR97, BM02, BP06, BP+11a}. Specifically, clouds should
have a minimum column density to be detected; and if the column
density where too large, clouds will become optically thick (in
\ala{12}CO), and then such dense regions would not be easily detected.
The net result of these effects would be that observed CO clouds
should exhibit a small dynamic range in column density, independent
of their intrinsic structure.

In a recent study, \citet{Lombardi+10} used near-infrared extinction
measurements to increase the dynamic range of inferred mass column
densities.  They studied a sample of local MCs with substantially
different physical properties, from giant MCs hosting massive star
formation, and masses of the order of 2$\times 10^5$ \Msun (e.g.,
Orion), down to clouds with less star formation activity, and masses
of the order of several thousand \Msun\ (the Pipe Nebula). Their main
results are:

\begin{itemize} 

 \item{} For an ensemble of MCs observed at a given extinction
   threshold $A_0$, the mass varies with size as $M\propto R^2$,
   implying that the column density $\Sigma = N/\mu m_H$ is constant
   and depends on that threshold.

\item{} The numerical value of the column density changes with the
  value of the threshold $A_0$.

\item{} {When analyzing a single MC at different thresholds $A_0$, the
  mass-size relation varies as a power-law, with a somewhat shallower
  power law index ($M\propto r^{1.6}$). }

\end{itemize}

\citet{Lombardi+10} argue that their results can be understood as a
consequence of the fact that clouds exhibit a column density
probability distribution function (\Npdf) with a lognormal functional
form, although in a recent paper, \citet{Beaumont+12} revisit the
\citet{Lombardi+10} analysis, and suggest (without demonstration) that
other forms than a lognormal \Npdf\ can produce the same result.  They
also argue that the only requirement to satisfy Larson's third
relation is that the mean \Npdf\ is uncorrelated with region-to-region
dispersion in area.

In the present work we discuss why MCs with very different properties
appear to have very similar column densities at a given threshold.  In
\S\ref{sec:toy} we draw \SigmaA\ diagrams to show that the main
features found by \citet{Lombardi+10} are found for very different
functional forms, and that the scatter in column density at any
threshold is smaller than a factor of 3.  In \S\ref{sec:thresholding}
we show, furthermore, that when thresholding the volume density, what
appears to be constant is the volume density, suggesting that the
relation is a consequence of the procedure adopted. We argue that the
only special feature of the structure of local MCs is that they peak
at a particular value of the mean \Npdf, in agreement with
\citet{Beaumont+12}.  {In \S\ref{sec:intracloud}, we analyze the
  case of the mass-size relation for a single cloud, showing that
  rather than a power-law, the relation is a curve that has slope
  approaching to 2 at small radii, and flattens for larger radius. As a
  consequence, there is some interval in which the slope is close to
  1.6-1.7, as reported in observations.  The properties of this
  relation are determined directly by the shape of the \Npdf.}  In
\S\ref{sec:discussion} we further identify the physical reasons for
having a typical \Npdf: First, there is a minimum column density for
CO shielding; and second, above this threshold the amount of mass
decreases rapidly with increasing column density, which is probably
the result of gravitational contraction and collapse.

\section{Formalism: Column density PDFs and their \SigmaA\ diagram} 
\label{sec:toy} 

In our analysis we follow the formalism outlined by
\citet{Lombardi+10}. We use $A$ to denote the
extinction in the $K$ band unless otherwise specified. We also recall
that $A_V \sim 10 A_K$, and that the \Npdf\ published by
\citet{Kainulainen+09} have a peak at $A_k\sim$~0.1, equivalent to
$A_V\sim 1$ or $N\sim 10^{21}$~cm\alamenos 2.  Now, given a
probability distribution function of the extinction $p(A) dA$, the
area and mass above some extinction threshold $A_0$ are defined as:

\begin{equation}
  S(A_0) = S_{\rm tot}\ \int_{A_0}^\infty \ p(A)\ dA
\label{eq:surface}
\end{equation}
and

\begin{equation}
  M(A_0) = S_{\rm tot} \mu m_H\ \beta \int_{A_0}^\infty A\ p(A)\ dA ,
\label{eq:masa}
\end{equation}
where $\mu$ is the mean molecular weight within the MC, $m_H$ is the
mass of the hydrogen atom, and $\beta$ is the ratio between the total
hydrogen column density and the dust extinction coefficient.
\citet[][see their Fig 4]{Lombardi+10} showed that if clouds have a
lognormal \Npdf, the surface density as a function of $A_0$, which is
given by the ratio of eq. (\ref{eq:masa}) divided by eq.
(\ref{eq:surface}), is nearly constant for low thresholds $A_0$, and
varies approximately linearly for large thresholds, i.e.,

\begin{eqnarray}
  \Sigma = \ \left\{ \begin{array} {rl}
    \to    {\rm constant} & \mbox{if   $A_0 << A_{\rm max} $,}  \\
     \propto  A_0    &\mbox{if  $ A_0 >> A_{\rm max}$,} 
     \end{array} \right .
\label{eq:behavior}
\end{eqnarray}
where $A_{\rm max}$ is the extinction value at which the \Npdf\ peaks. Since
this behavior reproduces qualitatively that of the 12 MCs in the Solar
neighborhood they analyzed, these authors suggest that the observed
constancy of the column density for different clouds is due to their
intrinsic internal structure, given by a lognormal \Npdf.  However,
many of the MCs in the solar neighborhood do not exhibit a lognormal
\Npdf\ at large column densities but rather frequently exhibit
power-law or more complicated shapes
\citep{Froebrich+07}. Thus, the behavior of $\Sigma(A_0)$ described
above cannot be consequence of a universal distribution function of
the column density of the MCs.

In order to understand the origin of Larson's third relation for
multiple clouds, we analyze the following examples. For clarity's
sake, we will first describe the examples, leaving the discussion for
\S\ref{sec:Concl_SigmaA}.

\section{The Inter-cloud mass-size relation for simple column density pdfs}

\subsection{A single power-law}\label{sec:singlepowerlaw}

First, assume that the \Npdf\ has a power-law
shape\footnote{\label{notaalpie} Note that plotting vs. $d\log A$
  will produce a power law with a different index,

\begin{eqnarray}
  m&=&n-1 .  \nonumber
\end{eqnarray}}:

\begin{equation}
  p(A) dA = p_0 (A/A_{\rm max})^{-n} dA
\label{eq:pdf_powerlaw}
\end{equation}
Using this equation, the mass above some extinction threshold is given
by

\begin{equation}
  M(A_0) = -{\Stotmath \mu m_H \beta p_0 A_{\rm max}^2\over 2-n}
  \biggl({A_0\over A_{\rm max}}\biggr)^{2-n} ,
\label{eq:mass_singlepowerlaw}
\end{equation}
were we have assumed that $n>2$ in order to get a finite mass
\citep[indeed, by simple inspection one can verify that all the column
  density histograms of MCs published in the literature {fall
    faster than a power-law with an exponent of $m\ge 2.5$, i.e.,
    $n\ge 3.5$, see e.g.,} ][]{Kainulainen+09}.  Similarly, the area
of the cloud will be given by

\begin{equation}
  S(A_0) = - {\Stotmath p_0 A_{\rm max}\over 1-n} \biggl({A_0\over
    A_{\rm max}}\biggr)^{1-n}
\label{eq:surface_singlepowerlaw}
\end{equation}
and thus, the column density will be simply given by

\begin{equation}
  \Sigma \equiv {M(A_0)\over S(A_0)} = \mu m_H \beta A_{\rm max} \biggl( {1-n
    \over 2-n}\biggr) \biggl({A_0\over A_{\rm max}}\biggr)  .
  \label{eq:Sigma_onepowerlaw}
\end{equation}
We note then, that for a power-law distribution, the column density
always varies linearly with $A_0$ \citep[see also][]{Beaumont+12}.

\subsection{Two power laws}\label{sec:two_powerlaws}

Now assume that the \Npdf\ has a two power-law shape: one with a
positive exponent, for low extinctions, and the other with a negative
exponent, for large extinctions, i.e.,

\begin{eqnarray}
  p(A) = \left\{ \begin{array} {rl}
      p_0 (A/A_{\rm max})^{q}  & \mbox{for $A\le A_{\rm max}$,} \\ 
      p_0 (A/A_{\rm max})^{-n} & \mbox{for $A\ge A_{\rm max}$,}
\end{array} \right . 
\label{eq:pdf_2powerlaw}
\end{eqnarray}
with $q, n > 0$.  Plugging this \Npdf\ into eqs.  (\ref{eq:surface})
and (\ref{eq:masa}) we obtain

\begin{equation}
  S = K_1 \biggl\{ {1\over 1+q}\biggl[ 1-\biggl({A_0\over
      A_{\rm max}}\biggr)^{1+q} \biggr] - {1\over 1-n}\biggr\}
\label{eq:S2powerlaw}
\end{equation}
with

\begin{equation}
  K_1 =\Stotmath p_0 A_{\rm max} 
\end{equation}
and

\begin{equation}
  M = K_2 \biggl\{ {1\over 2+q}\biggl[
    1-\biggl({A_0\over A_{\rm max}} \biggr)^{2+q} \biggr] - {1\over
    2-n}\biggr\}
\label{eq:M2powerlaw}
\end{equation}
with

\begin{equation}
  K_2= \Stotmath \mu m_H \beta p_0 A_{\rm max}^2 .
\end{equation}
Making $K_3=K_2/K_1 = \mu m_H \beta A_{\rm max}$, the surface density
is given by

\begin{equation}
  \Sigma(A_0) = K_3 { \biggl[ 1-\bigl({A_0/
        A_{\rm max}}\bigr)^{2+q} \biggr]{(2+q)^{-1}} - (2-n)^{-1} \over
    \biggl[ 1 - \bigl({A_0/A_{\rm max}}\bigr)^{1+q} \biggr]{(1+q)^{-1}} -
    (1-n)^{-1} }
    \label{eq:Sigma_twopowerlaw}
\end{equation}
for $A_0 \le A_{\rm max}$. For larger $A_0$, the case is reduced to a
single power-law, and the surface density will be given by eq.
(\ref{eq:Sigma_onepowerlaw}).

\begin{figure}
\includegraphics[width=1.\hsize]{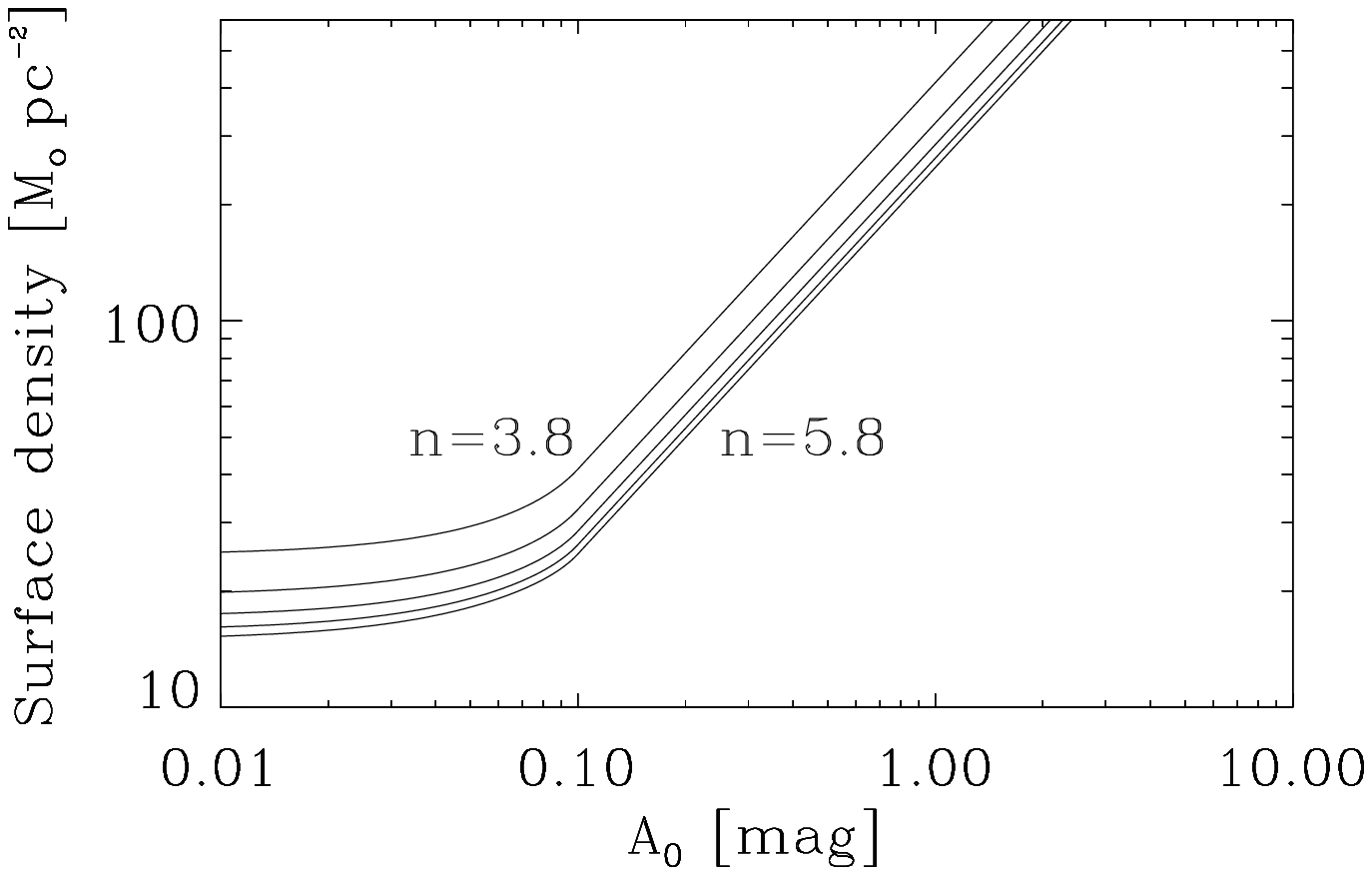}
\includegraphics[width=1.\hsize]{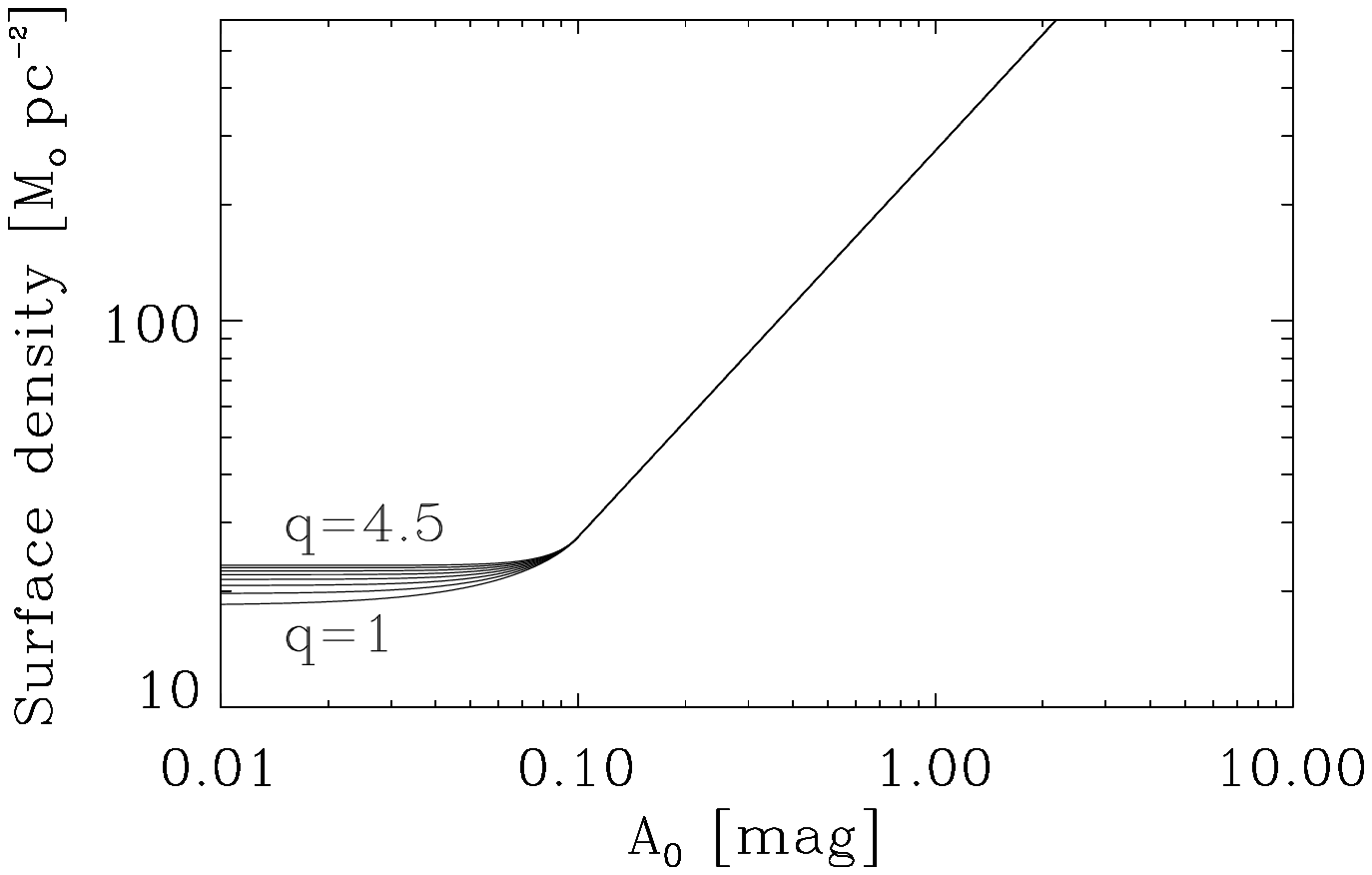}
\caption{\label{fig:Sigma_twopowerlaw} $\Sigma$ vs $A_0$, from cloud
  \Npdf s given by two power laws, as in eq. (\ref{eq:pdf_2powerlaw}).
  The upper panel presents curves with $q=0$ and $n=2.8$, 3.3, 3.8,
  4.3, and 4.8. Lower panel: curves with $n=$, and $q=1$ to 5.5 in
  steps of 0.5. Note that in both cases, the inferred column density
  $\Sigma$ is the same, within a factor of $\sim$2.}
\end{figure}

In Fig.~\ref{fig:Sigma_twopowerlaw} we plot $\Sigma$ vs $A_0$ for a
two power-law \Npdf.  In both cases, we have assumed $A_{\rm
  max}=0.1$, the typical value found in a large sample of Solar
neighborhood MCs by \citet{Kainulainen+09}. In the upper panel we vary
{$n$ from $\sim 3.8$ to 5.8, in steps of 0.5. This is
  approximately the range of exponents that can be found in the \Npdf
  s presented by \citet[][$m\sim 2.8$ to 4.8.]{Kainulainen+09} } As
expected, the change in the power-law index modifies the surface
density at every $A_0$.  In the lower panel, however, varying the
low-extinction index $q$ (from 0 to 4.5 in steps of 0.5) only modifies
the surface density for low $A_0$.

Two more features can be inferred at this point. First of all, for
large $A_0$, this case is reduced to the previous one (a single
power-law), and then, the column density varies linearly with the
extinction threshold, i.e., $\Sigma\propto A_0$. On the other hand, at
low $A_0$ ($<< A_{\rm max}$), the column density for a cloud with a
given \Npdf\ becomes constant:

\begin{equation}
  \Sigma(A_0) \simeq \mu m_H \beta A_{\rm max} \biggl( {1+q\over 2+q}
  \biggr) \biggr( {1-n\over 2-n} \biggr).
\label{eq:masa4powerlaw}
\end{equation}
We notice, moreover, that for typical values\footnote{Typically
  \citep[see ][]{Kainulainen+09}, $3.8 \le n \le 5.5$ and $-1 \le q\le
  5$ (recall that $q=1$ corresponds to a flat \Npdf, according to
  footnote~\ref{notaalpie}), where $m$ is the slope measured in the
  \Npdf\, and $q$ plays the role of $n$.} of $n, q$, the minimum
column density given by eq. (\ref{eq:masa4powerlaw}) varies less than
a factor of two.

\subsection{A lognormal \Npdf}

The case of a lognormal \Npdf\ was analyzed by \citet{Lombardi+10},
and the mass surface density is given by

\begin{eqnarray}
  \Sigma(A_0) &=& \mu m_H \beta A_{\rm max} \exp\biggl({\sigma^2\over
    2}\biggr) \times \nonumber \\ 
              & & {1-{\rm erf} \{ [\ln{(A_0/A_{\rm max})}
      - \sigma^2]/\sqrt{2}\sigma \}\over 1-{\rm erf} [\ln{(A_0/A_{\rm max})}
      /\sqrt{2} \sigma] } .
    \label{eq:Sigma_lognormal}
\end{eqnarray}

Although the non-dimensional plot of this function has been already
presented and discussed by these authors, for comparison
we display it in dimensional form, assuming $A_{\rm max} = 0.1$. 
The qualitative behavior is similar to that of the
\Npdf\ exhibiting two power-laws previously discussed.

\begin{figure}
\includegraphics[width=1.\hsize]{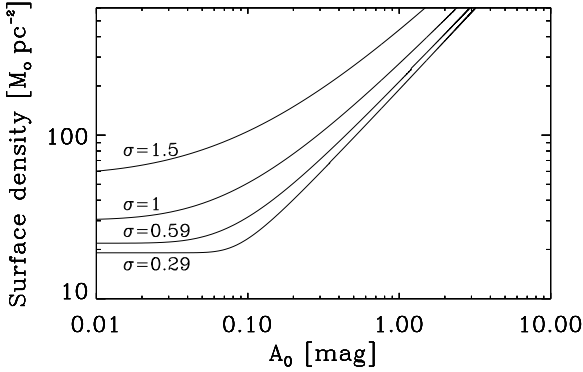}
\caption{\label{fig:Sigma_lognormal} $\Sigma$ vs $A_0$, for a
  cloud that exhibits a \Npdf\ given by a lognormal function. Here we
  assume $A_{\rm max}=0.1$. The different curves follow
  eq.~(\ref{eq:Sigma_lognormal}) with $\sigma = 0.29$ and 0.59
  \citep[the range of variation in $\sigma$ for the lognormal
    functions fitted by][]{Kainulainen+09}, plus two extreme cases:
  $\sigma =1$, and 1.5.  We note that, even for these unrealistic
  cases, the inferred column density $\Sigma$ is the quite the same,
  within a factor of $\leq$ 3--5.}
\end{figure}

We further note that, since the error function can be expanded as

\begin{equation}
{\rm erf}(x) \simeq 1 - {\exp{(-x^2)}\over \sqrt\pi} \biggl(x^{-1} -
{1\over 2}x^{-3}+{3\over 4} x^{-5}+ ... \biggr) ,
\end{equation}
to first order the mass column density can be written as:

\begin{eqnarray}
  \Sigma & = & \mu m_H \beta A_{\rm max} \exp{\biggl({\sigma^2\over 2}
    \biggr)} \ln (A_0/A_{\rm max}) \times \nonumber \\ 
         &  &  \biggl({\exp{ [-{
          (\ln({A_0/A_{\rm max}}) - \sigma^2)^2 / \sqrt 2 \sigma} ] } \over
    (\ln({A_0/ A_{\rm max}}) - \sigma^2) \exp{[-\ln({A_0/
        A_{\rm max}})/2\sigma^2]} }\biggr) , \nonumber
\end{eqnarray}
for large values of $A_0$.  Simplifying,

\begin{equation}
  \Sigma = \mu m_H \beta A_{\rm max} {({A_0/A_{\rm max}}) \ln{(A_0/A_{\rm max})} \over
    \ln{(A_0/A_{\rm max})} - \sigma^2 } .
\end{equation} 
This equation makes clear that, for large $A_0/A_{\rm max}$, $\ln(A_0/A_{\rm max}) >>
\sigma^2$ and thus, 

\begin{equation}
  \Sigma\propto A_0 .
\end{equation}
On the other hand, for small $A_0$, $\Sigma$ becomes constant with a
value of

\begin{equation}
  \Sigma(A_0) = \mu m_H \beta A_{\rm max} \exp{\sigma^2\over 2} .
\end{equation}

\subsection{Column density PDF  with several peaks}\label{sec:complex}

We now proceed with a slightly more complicated example: a
three-peaked \Npdf\ defined by three lognormal functions, as follows.

\begin{equation}
  p(A) dA = \sum_{i=1, 3} {p_i\over A} \exp{\biggl(-{(\ln A - \ln
      A_i)^2\over 2\sigma_i^2} \biggr)} dA ,
\end{equation}
and thus, the surface density will be given by

\begin{eqnarray}
  \Sigma & = & {\mu m_H \beta\over \sqrt{2\pi} } \times \biggl\{ \nonumber \\ 
         & & {\sum_{i=1,3} p_i A_i e^{\sigma_i^2/2} \bigl[ 1-{\rm erf}
                 \bigl( (\sigma_i - \ln{(A_0/A_i)} )/\sqrt 2\sigma_i
                 \bigr) \bigr] }/ \nonumber \\ 
         & & {\sum_{i=1,3} p_i \sigma_i \bigl[ 1-{\rm erf}
                 [\ln{(A_0/A_i)}/ \sqrt \sigma_i ] \bigr] } \biggr\}
\end{eqnarray}

\begin{figure}
\includegraphics[width=1\hsize]{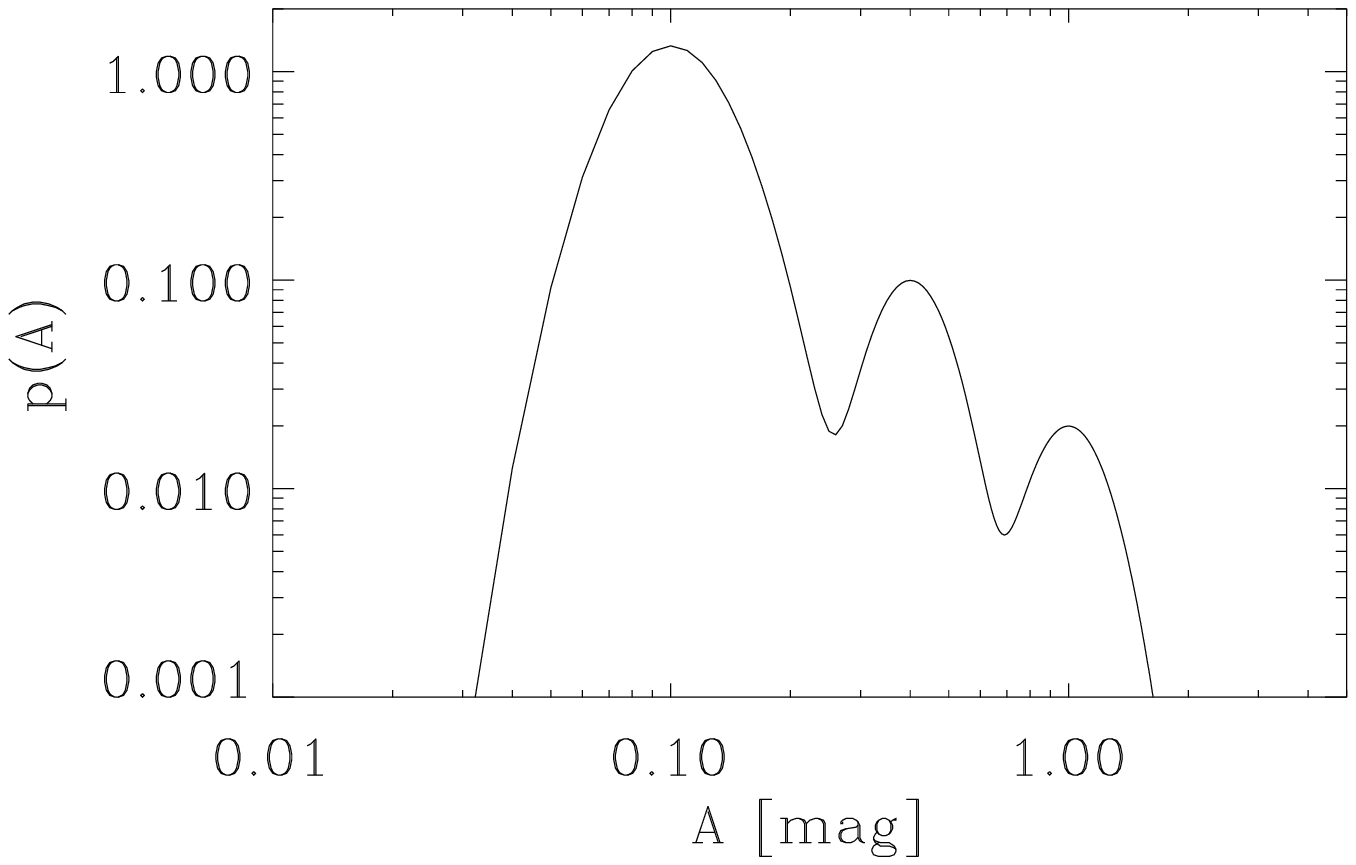}
\includegraphics[width=1.\hsize]{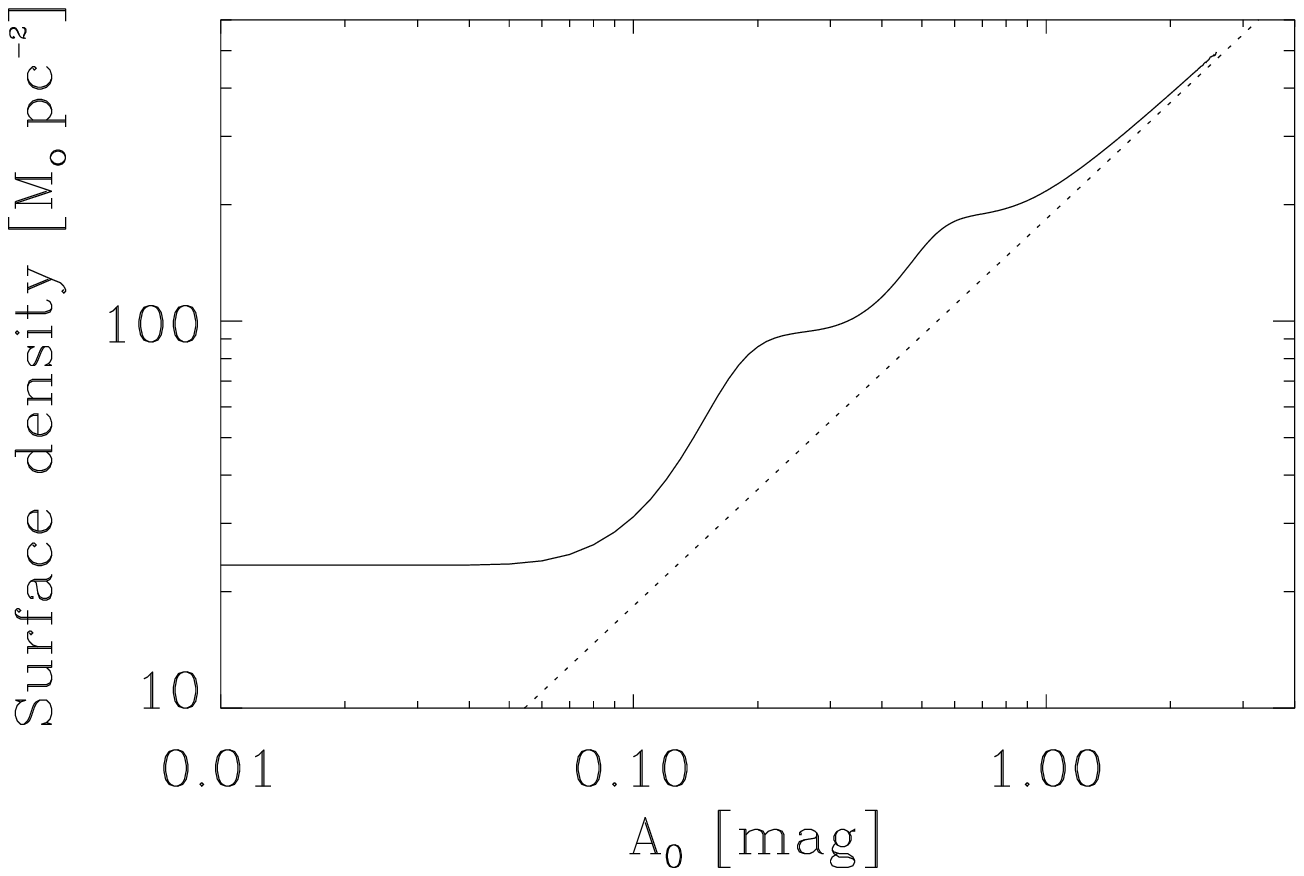}
\caption{\label{fig:3lognormales} {\it Upper panel:} \Npdf\ of a
  hypothetical MC exhibiting three peaks. {\it Lower panel:} its
  corresponding $\Sigma$ vs $A_0$ diagram. The dotted line denotes the
  line with slope $\beta_K^{-1}$, where $\beta=1.67\times
  10^{22}$cm\alamenos 2 mag\alamenos 1, the opacity in the $K$ band.}
\end{figure}

In Fig~\ref{fig:3lognormales} we show our hypothetical \Npdf\ (upper
panel), and its respective $\Sigma$ vs $A_0$ function (lower
panel). The parameters of the lognormal functions were chosen to
enhance their effect on the column density. In reality, the observed
\Npdf s exhibit smaller variations.  As can be noted from this figure,
we again reproduce the main features observed in the \SigmaA\ diagram:
for $A_0 >> A_1$, $\Sigma$ increases linearly with $A_0$, and for for
$A_0 << A_1$, $\Sigma$ becomes constant.  Additionally, in this case
we note that the peaks in the \Npdf\ produce small fluctuations with a
trend that has a slope close, but smaller than unity. We will get back
to such fluctuations below.

\subsection{Lognormal with a power-law tail PDF}\label{sec:log+power}

\begin{figure}
\includegraphics[width=1.\hsize]{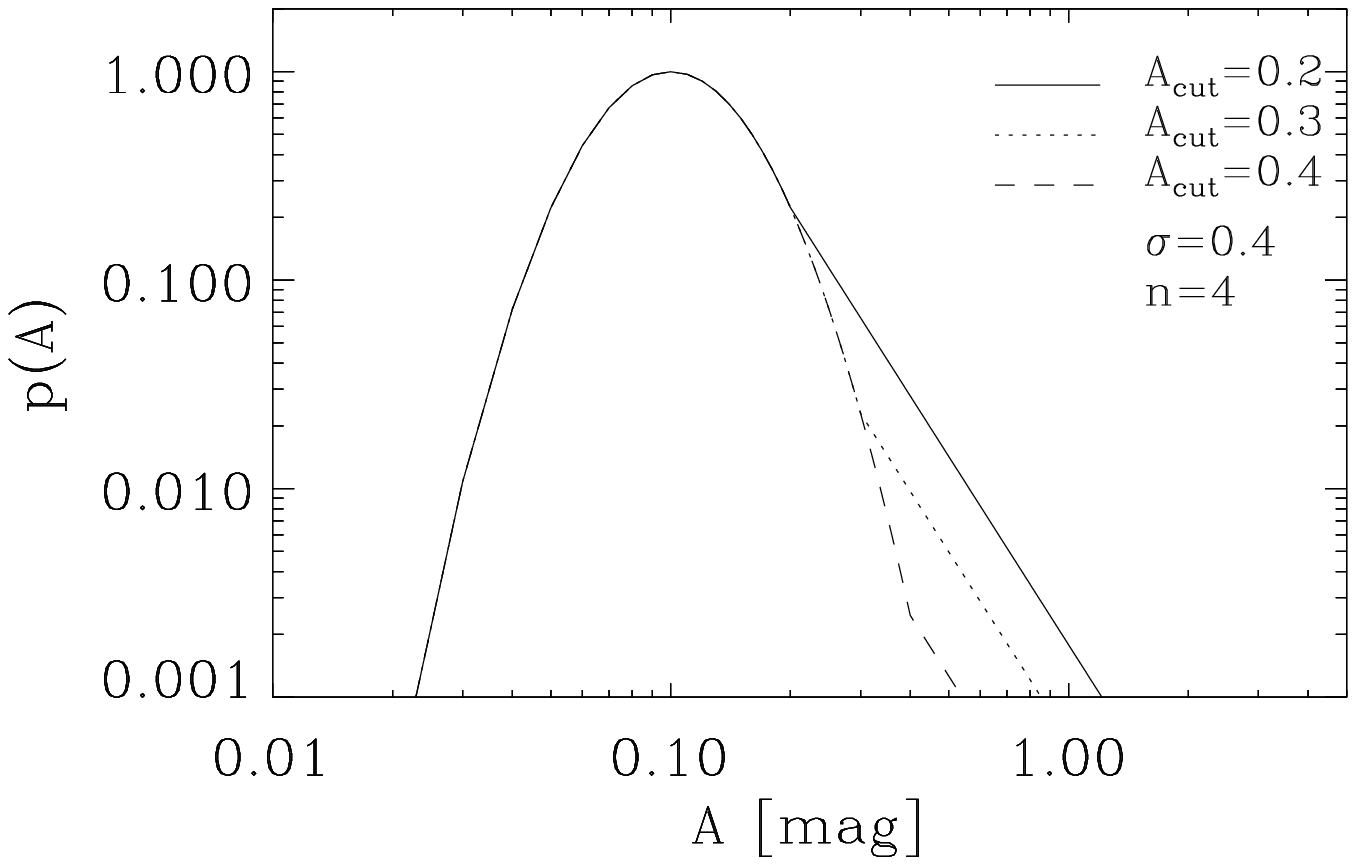}
\includegraphics[width=1.\hsize]{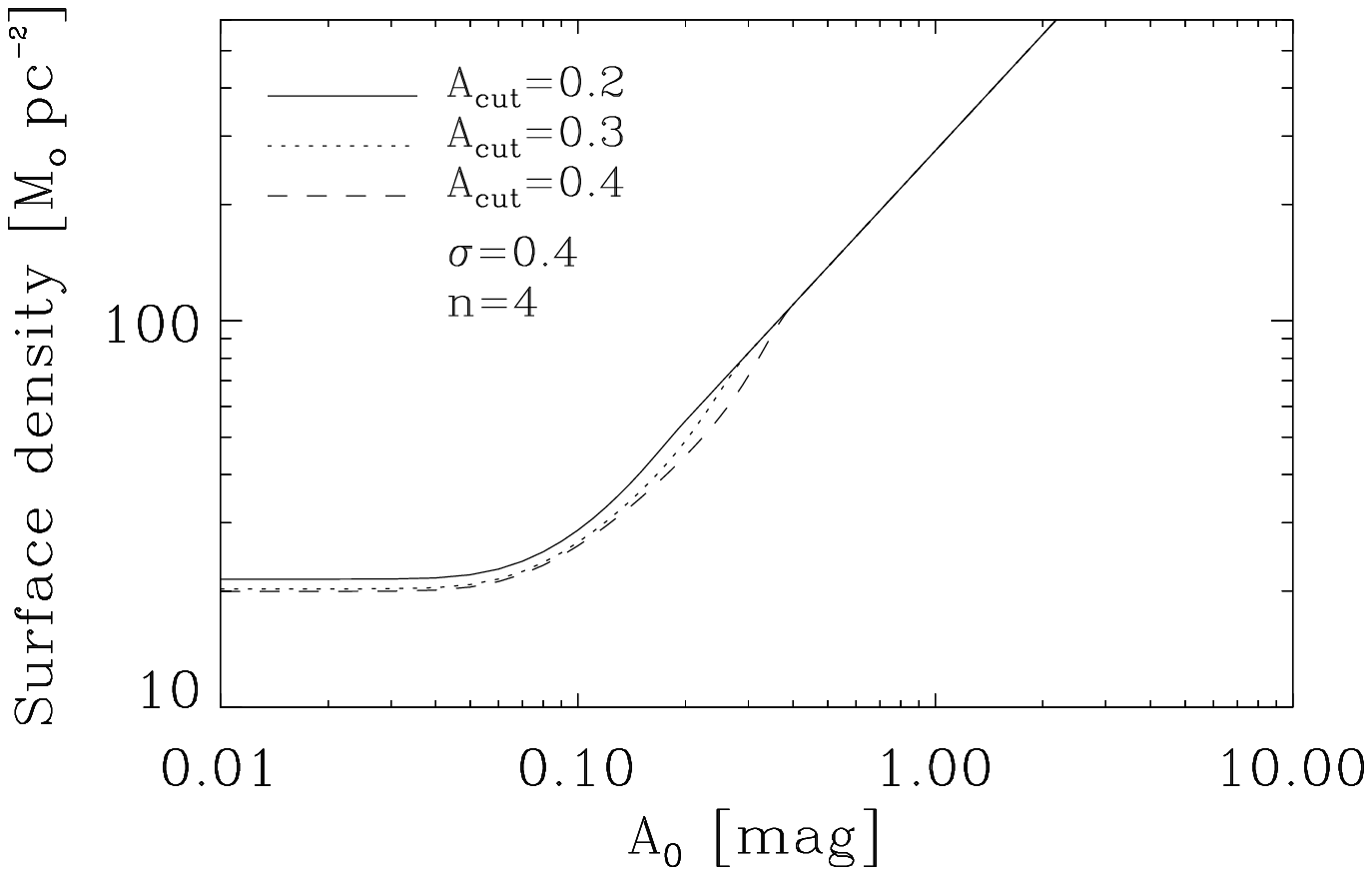}
\caption{\label{fig:log_power_A2} {\it Upper panel:} \Npdf\ defined by
  a lognormal function at low extinctions and a power-law tail at large
  extinctions, for different values of $A_{\rm cut}$: 0.2, 0.3, and
  0.4. $\Sigma=0.4$ and $n=4$ ($m=3$) remain fixed. {\it Lower panel:}
  The corresponding \SigmaA\ diagram. Notice that in the power-law
  regime ($A>A_{\rm cut}$), the column density is the same for the
  three cases, since they have the same value of $n$ and $A_{\rm
    cut}$, (see eq. (\ref{eq:Sigma_onepowerlaw})).}
\end{figure}

\begin{figure}
\includegraphics[width=1.\hsize]{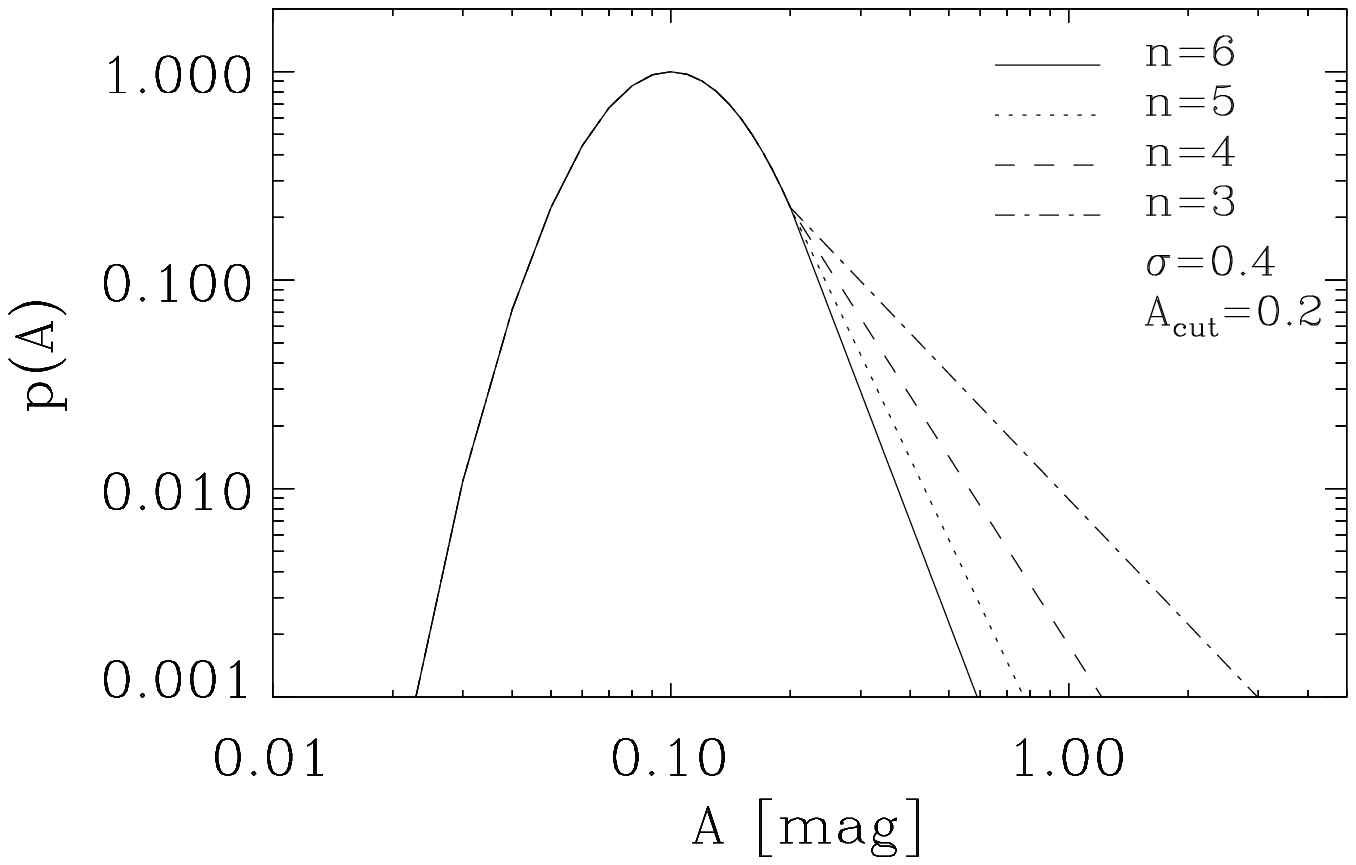}
\includegraphics[width=1.\hsize]{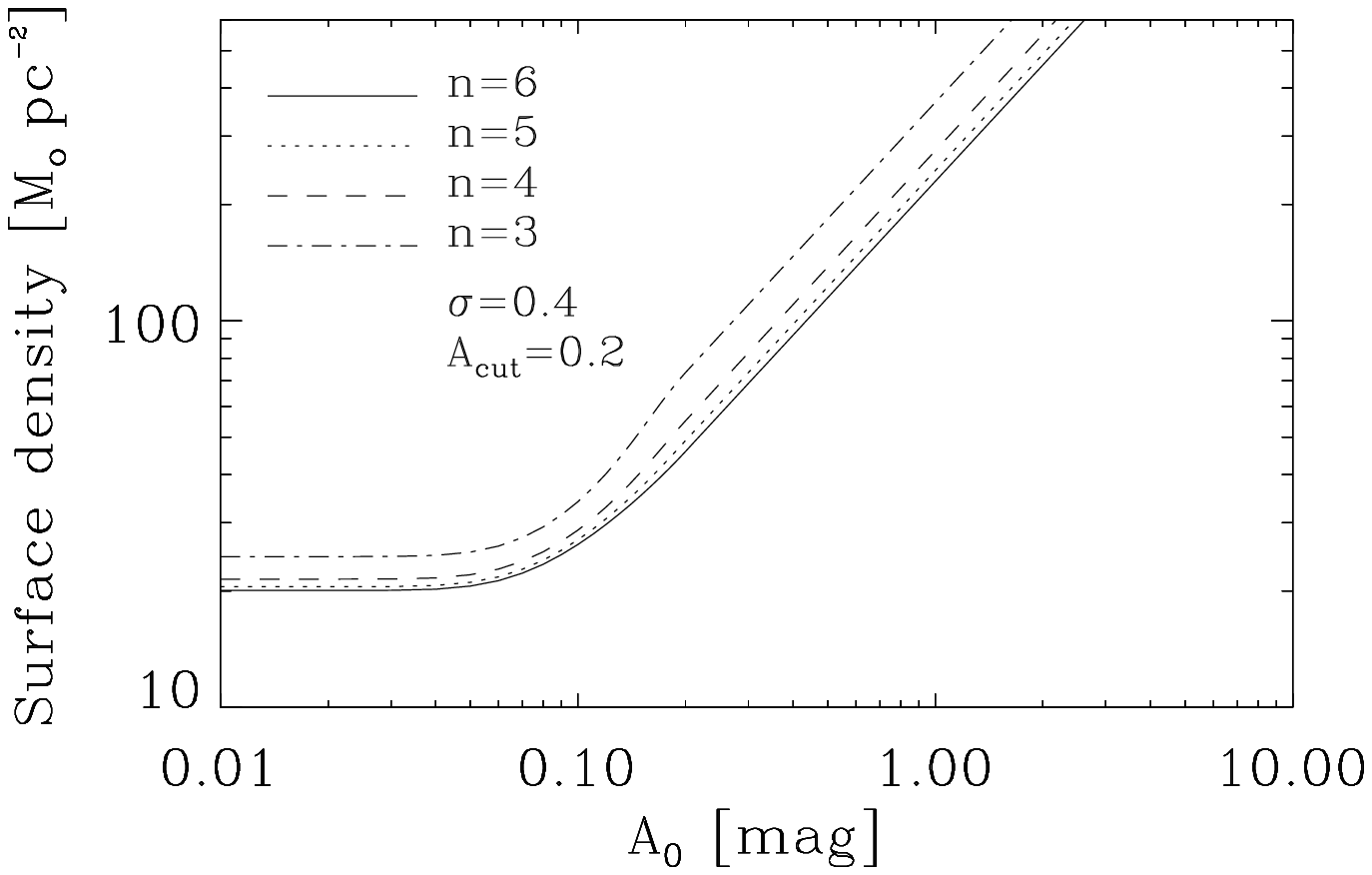}
\caption{\label{fig:log_power_n} {\it Upper panel:} \Npdf\ defined by
  a lognormal function at low extinctions and a power-law tail at
  large extinctions, for different values of $n$: 3, 4, 5, and
  6. $\Sigma=0.4$ and $A_2=0.2$ remain fixed. Note that the slope that
  one measures in this plot is $m=n-1$.  {\it Lower panel:} The
  corresponding \SigmaA\ diagram. }
\end{figure}

\begin{figure}
\includegraphics[width=1.\hsize]{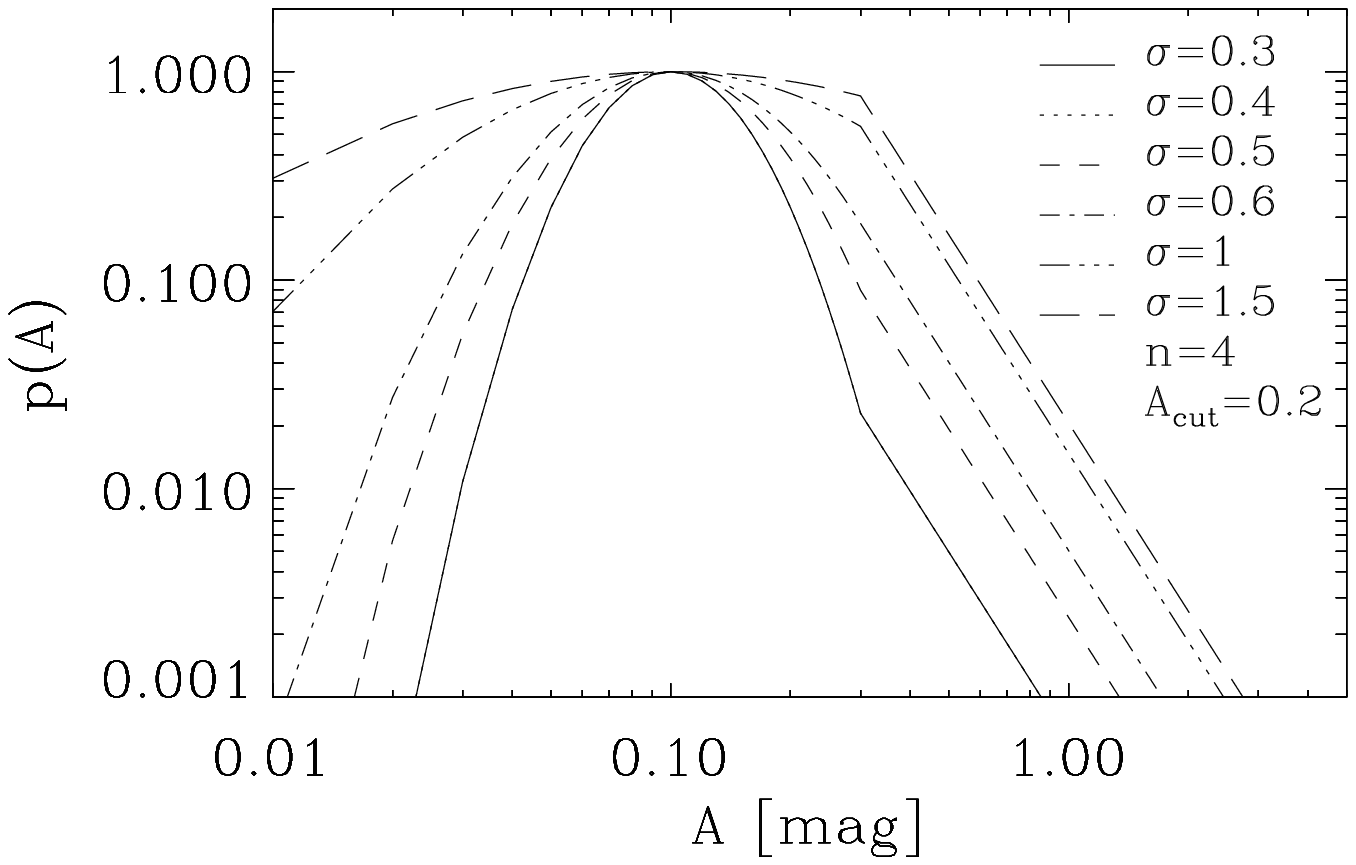}
\includegraphics[width=1.\hsize]{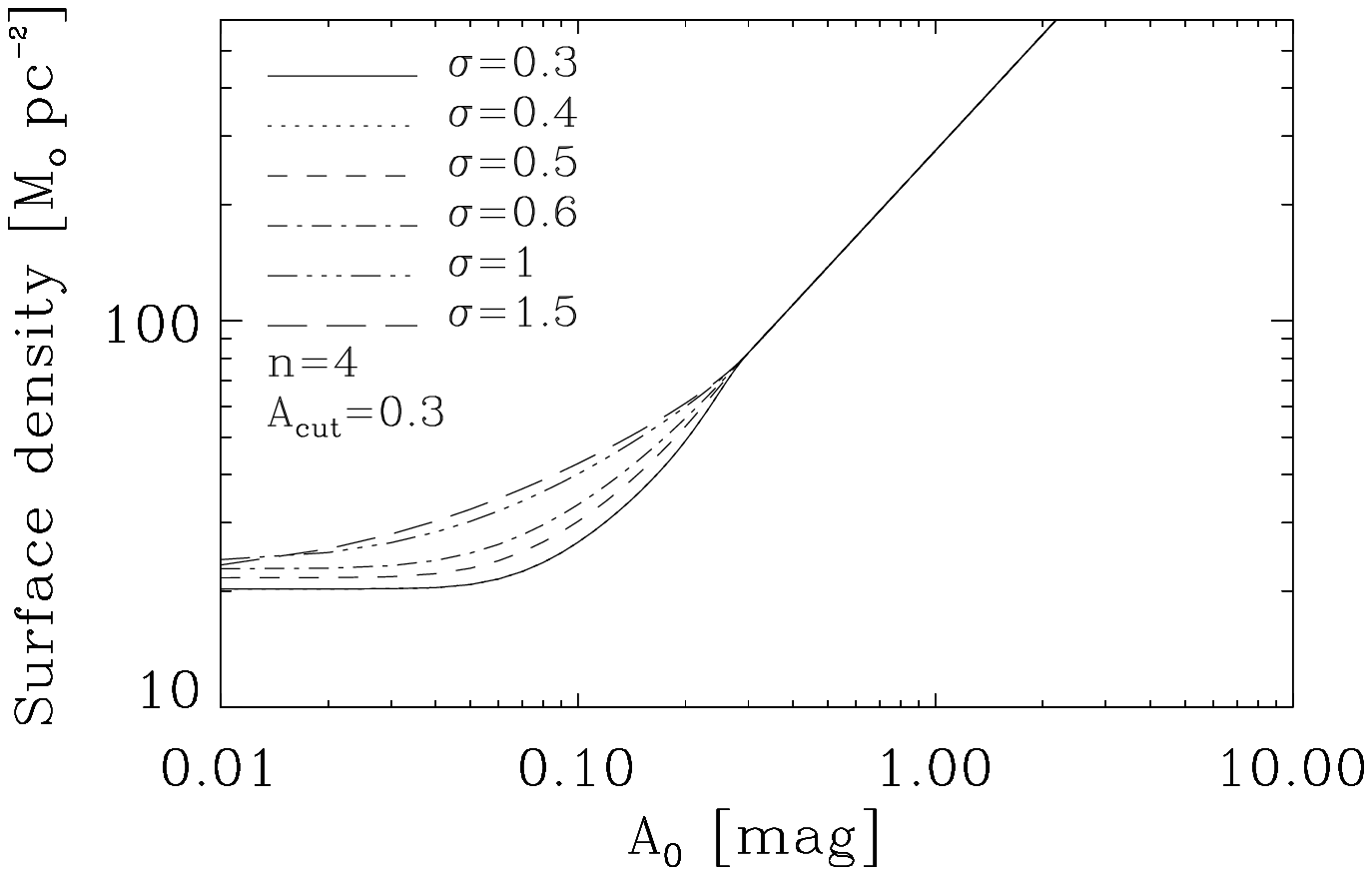}
\caption{\label{fig:log_power_s1} {\it Upper panel:} \Npdf\ defined by
  a lognormal function at low extinctions and a power-law tail at
  large extinctions, for different values of $\sigma$: 0.3, 0.4, 0.5,
  0.6, and the unrealistic cases with $\sigma=$1 and 1.5. $A_2=0.3$
  and $n=4$ ($m=3$) remain fixed. {\it Lower panel:} The corresponding
  \SigmaA\ diagram. As in the case of Fig.~\ref{fig:log_power_A2}, the
  column density in the power-law regime ($A>A_{\rm cut}$) is the same
  for the three cases, since they have the same value of $n$ and
  $A_{\rm cut}$.}
\end{figure}

Most of the observed column density histograms appear to behave as a
lognormal function, at low extinctions, with a power-law tail, at
larger $A$ \citep[e.g., ][see also Ballesteros-Paredes et al. 2011b
  for the theoretical counterpart]{Kainulainen+09}.  The functional
form of such \Npdf\ is, thus,

\begin{eqnarray}
  p(A) dA = \left\{ \begin{array} {rl}
           p_0 \exp{\biggl(-{(\ln A - \ln
           A_{\rm max})^2\over 2\sigma^2} \biggr)} dA/A  
                & \mbox{if $A \le A_{\rm cut} $,} \\
            p_1({A/A_{\rm cut}})^{-n} dA
                & \mbox{if $A \ge A_{\rm cut} $,} 
                \end{array} \right .
\label{eq:Npdf_lognormal+powerlaw}
\end{eqnarray}
where $\ln{A_{\rm max}}$ denotes the extinction value at which the
lognormal is maximum, and $\ln{A_{\rm cut}}$ the place where the
\Npdf\ changes from a lognormal to a power-law function.  The
continuity condition at $A_{\rm cut}$ is:

\begin{equation}
  p_1 = \biggr({p_0\over A_{\rm cut}}\biggr) \exp\biggl(-{(\ln A_{\rm
      cut}-\ln A_{\rm max})^2\over 2\sigma^2}\biggr) .
\label{eq:continuidad}
\end{equation}
As in the case of two power-laws (\S\ref{sec:two_powerlaws}), at large
extinctions ($A>A_2$), the mass, surface and column density must be
given by the solution of a single powerlaw, eqs.
(\ref{eq:mass_singlepowerlaw}), (\ref{eq:surface_singlepowerlaw}) and
(\ref{eq:Sigma_onepowerlaw}).  For smaller extinctions ($A< A_{\rm
  cut}$), however, the solutions for $M$ and $S$ are:

\begin{eqnarray}
  S(A_0) = S_{\rm tot} \biggl\{  p_0 \sqrt{\pi\over 2}
  \sigma\biggl[ {\rm erf}\biggl({\ln(A_{\rm
        cut}/A_{\rm max})\over\sqrt{2}\sigma}\biggr) -  \nonumber \\
    {\rm erf}\biggl({\ln(A_0/A_{\rm max})\over \sqrt{2}\sigma }\biggr) \biggr]
    - {p_1 A_{\rm cut}\over 1-n} \biggr\} ,
\label{eq:surface_lognormal+powerlaw}
\end{eqnarray}
and

\begin{eqnarray}
  M(A_0) = S_{\rm tot} \mu m_H \beta\ \biggl\{ - p_0 \sqrt{\pi\over 2}
  \sigma {\rm e}^{\sigma^2/2} A_{\rm max}  \nonumber \\ 
\biggl[ {\rm erf}\biggl({\sigma^2-\ln A_{\rm cut}+\ln{A_{\rm
      max}}\over\sqrt{2}\sigma }\biggr) - \nonumber \\
{\rm erf}\biggl({\sigma^2-\ln A_0 + \ln A_{\rm max}\over
  \sqrt{2}\sigma }\biggr) \biggr] - 
\nonumber\\ {p_1 A^2_{\rm
      cut}\over 2-n} \biggr\} ,
\label{eq:mass_lognormal+powerlaw}
\end{eqnarray}
while the column density is given by the ratio of eqs.
(\ref{eq:mass_lognormal+powerlaw}) over
(\ref{eq:surface_lognormal+powerlaw}).  We now plot the \Npdf\ as a
function of the extinction $A$ (upper panels), and its corresponding
mass column density $\Sigma$ as a function of the extinction threshold
$A_0$ (lower panels).  In Fig.~\ref{fig:log_power_A2} we vary the
value of $A_{\rm cut}$, the extinction value where we the power-law
cuts the lognormal. In fig.~\ref{fig:log_power_n} we vary the value of
the power-law index $n$, while in Fig.~\ref{fig:log_power_s1} we vary
the value of the standard deviation $\sigma$ of the lognormal
function.

As we can see, again, the column density is basically the same in all
cases, regardless of the parameter we are varying, and large
differences occur only when we strongly vary the power-law index $n$.

\subsection{Conclusions from the \SigmaA\ diagrams}\label{sec:Concl_SigmaA}

In the previous sections we have constructed different functional
forms of the \Npdf. Their main features are: (i) they peak at the same
value ($A_1\sim 0.1$), (ii) and they decrease strongly for larger
extinction values. In all cases we have recovered the main two
features observed for the column density of MCs: at low extinction
thresholds, the column density is nearly constant, while at large
extinction thresholds the column density increases linearly with the
threshold.  Additionally, in all cases the values obtained for
$\Sigma$ at any $A_0$ are quite similar, despite their strong
differences in functional form of the \Npdf.

The reason for which very different \Npdf s produce approximately the
same values of the observed column density at a given threshold,
appearing to reproduce the $M\propto R^2$ relation, is a consequence
of the definition of $\Sigma$ itself, along with the fact that the
\Npdf s of MCs decay fast: the column density given by the ratio of
eq. (\ref{eq:masa}) over eq. (\ref{eq:surface}) is the average of the
extinction coefficient (which is proportional to the mean surface
density) above the extinction threshold $A_0$.  Lets assume now that
the \Npdf\ has an arbitrary shape below its maximum at $A_1$, and
decreases strongly beyond that value.  For $A_0 < A_1$, the mean
column density does not change substantially with the threshold $A_0$
because the mass at a given extinction $A$, which is proportional to
the product of the number of points contributing to the \Npdf\ times
$A$, ($dm \propto A p dA$), will be dominated by the mass at the peak
$A_1$. Note that this is valid even if the \Npdf\ were flat for $A_0 <
A_1$. This is the case where the surface density is constant with
$A_0$ in the \SigmaA\ diagram.

As we approach the peak of the \Npdf, small changes in the threshold
$A_0$ start removing a substantial number of low extinction data
points from the \Npdf.  The calculation of our mean extinction starts
to be dominated by larger values of $A$, making the mean surface
density $\langle\Sigma\rangle$ to increase rapidly.  Once we have
passed the maximum $(A_0 > A_{\rm max})$, and the distribution decreases
rapidly, the mass will be dominated by values close -but above- to
$A_0$.  This is the range in which the $\langle\Sigma\rangle$ varies
linearly with the threshold, i.e., $\langle \Sigma\rangle\propto A_0$.

A similar analysis can be performed if the \Npdf\ exhibits several
bumps, as in the case of our three-peaked \Npdf\ (see
Fig.\ref{fig:3lognormales}): below each peak, the \SigmaA\ curve is
more or less constant because the mass is dominated by the next peak.
After each peak, the column density grows fast because we are removing
a substantial number of points at low column densities. Once we reach
the final peak, the column density should increase proportional to
$A_0$ because the mass is dominated by values slightly larger than
$A_0$.

We emphasize that the arguments given by \citet{Kegel89} and
\citet{Scalo90}, that clouds are not seen at large column densities
because of optically thick effects, do not apply because of the larger
dynamic range available in more recent observations.  Clouds
approximately fall along the constant column density line just because
we are measuring the mean value of a property which probability
distribution function has a peak at nearly the same value for all
objects, and falls fast enough.

On the other hand, from the present analysis it becomes clear how a
molecular cloud must behave in order to depart from the Larson's third
relation, i.e., in order to have, at a given threshold, a
substantially different column density compared to other clouds: (a)
it must have a wider \Npdf, or (b) it must peak at a substantially
different extinction value $A_{\rm max}$.  The first case can be
achieved by having either a \Npdf\ with a much flatter slope at large
column densities, or a wider lognormal. Nevertheless, all the \Npdf s
reported in the literature fall a factor of $\sim$ 3 orders of
magnitude when increasing one order of magnitude in column density
\citep{Kainulainen+09}, making this possibility inapplicable from the
observational point of view.  {Moreover, it has been shown
  theoretically that, for a supersonic turbulent field with Mach
  number $M$, the volume density PDF has a standard deviation
  $\sigma_\rho$ given by
 
\begin{equation}
   \sigma_{\log{\rho}} = \sqrt{\ln(1 + M^2/4)}
\label{eq:sigma_rho}
\end{equation}
\citep{Ostriker+99, PN02}.  Clearly, the dependence on $M$ is weak.
If one takes one of the most extreme cases known, the {G0.253$+$0.016}
molecular cloud near the galactic center \citep{Longmore+12}, with a
velocity dispersion of 16 \kms, the width of the {\em volume} pdf will
differ by only a factor of two from more typical clouds, like Taurus
with a dispersion of $\sim 2$ \kms.  But the column density PDF by
definition must have a smaller width, since the column density is the
integral of the volume density, and thus, density fluctuations are
smeared out.  Thus, one should not expect a wide lognormal \Npdf\ even
for large Mach numbers.}

\begin{figure}
\includegraphics[width=1.\hsize]{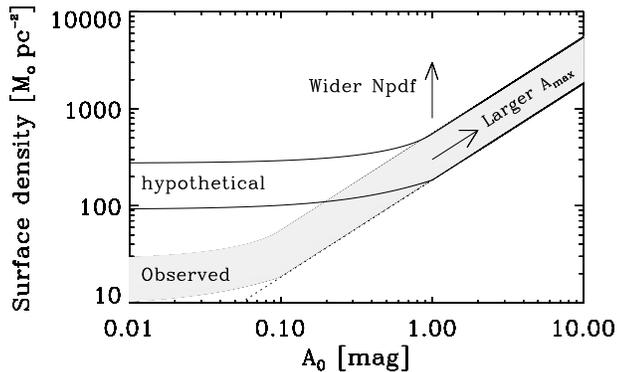}
\caption{\label{fig:irdc} Schematic \SigmaA\ diagram for local clouds
  (shadow region), and for some hypothetical clouds (white
  region). Note that, the wider the \Npdf, the wider the gray region.
  If the dust opacity of such hypothetical clouds is the same than
  that of the local clouds, their locus in this diagram should be
  shifted towards the upper right region, as indicated. Here we
  assumed that $A_{\rm max}\sim 1$ for these clouds. Observed
  \Npdf\ of IRDCs by \citet{Kainulainen+11} discard the possibility of
  these clouds.}
\end{figure}

In the second case, i.e., for hypothetical clouds peaking at
substantially different extinction values $A_{\rm max}$, {\it if} the
dust opacity $\beta$ is the same for such a cloud, its \SigmaA\ curve
will be shifted along the line with slope unity, increasing also its
column density for $A_0<A_{\rm max}$, as shown in Fig.~\ref{fig:irdc}.
From this figure, one may think that for extinction thresholds
$A_0<1$, such a cloud will exhibit a much larger column density at a
given threshold, compared to the Solar Neighborhood clouds.  {
  Although one may think that this could be the case of the infrared
  dark clouds (IRDCs), which achieve column densities substantially
  larger than local clouds with typical values of
  \diezala{22}~cm\alamenos 2 and maximum values up to
  \diezala{24}~cm\alamenos 2 \citep[e.g., ] [] {Ragan+11}, the
  cumulative mass functions of IRDCs \citet[][see their
    Fig.~7]{Kainulainen+11} show that the mass keeps increasing moving
  to lower extinction $A_V \sim 2$, indicating that the peak of their
  \Npdf\ must be at $A_v\sim 1-2$, just as local clouds.  }

\section{The effect of thresholding density}\label{sec:thresholding}

\begin{figure}
\includegraphics[width=1.\hsize]{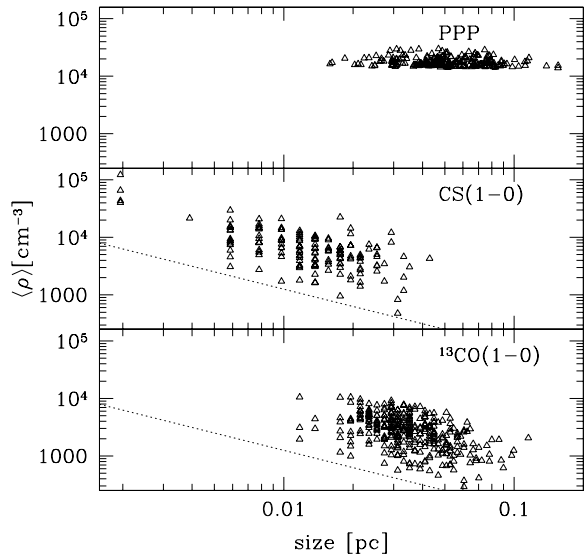}
\caption{\label{fig:BM02} Density-size relation for clumps in
  numerical simulations. Upper panel: real space, or
  position-position-position (PPP) space. The clumps are found by
  thresholding the volume density field.  Middle and lower panels:
  observational space, position-position-velocity (PPV) space. In this
  case, the clumps are found by thresholding the intensity, which in
  turn is proportional to the column density. Figure taken from
  \citet{BM02}.}
\end{figure}

Imagine now that we were able to measure the volume density fields of
MCs, and then threshold them in order to measure the mean volume
density of the clouds.  In this case, what we would obtain is a nearly
constant volume density, implying that $M\propto R^3$. To show that
this is the case, in Fig.~\ref{fig:BM02} we show the volume density
($y$ axis) against size ($x$ axis) for clumps in a numerical
simulation of isothermal molecular clouds with forced turbulence, as
published by \citet[][ Fig. 9]{BM02}.  The details of the particular
simulation used and on the assumptions used for the radiative transfer
performed to mimic the line emission can be found in that paper. Here
we stress that the same result can be found in simulations with or
without self-gravity, with or without magnetic fields, or whether the
turbulence is forced or decaying.  The data points in the upper panel
are obtained by thresholding the volume density, while the middle and
lower panels are obtained by thresholding the simulated CS(1-0) and
the \ala{13}CO(1-0) intensity, respectively, which is proportional to
the column density.  The dotted line in these panels has a slope of
$-1$, i.e., its a line with constant column density.

It can be seen from the upper panel that the volume density seems to
be constant for all cores, when we threshold the volume density.  When
we analyze the observational space (middle and lower panels), the
points follow the straight line with slope $-1$, implying that the column
density is constant.  In this case, the cores were found by
thresholding the intensity, which is proportional to the column
density. Thus, while we will deduce a relation $M \propto R^3$ from
the upper panel, from the lower two we will deduce that the relation
is $M\propto R^2$.

\section{The intra-cloud mass-size relation}\label{sec:intracloud}

We now turn to the intra-cloud mass-size relation, i.e., the $M-R$
relation for a single cloud.  In this case, Fig.~2 by
\citet{Lombardi+10} show that none of the $M-R$ diagrams exhibit a
power-law of the form\footnote{We use the letter $a$ to denote the
  exponent of the hypothetical $M-R$ power-law, and $\alpha$ to denote
  the slope of the curve of the $M-R$ curve, which is not a
  power-law.}:

\begin{equation}
M\propto R^{a}, 
\label{eq:MRsinglecloud}
\end{equation}
Instead, the $M-R$ diagram of all clouds is a curved line with the
following properties: (a) at the smaller radii, all curves exhibit a
power-law with slope of 2, and (b) for the larger radii, all the
curves become flatter, with slopes of the order of unity.  A natural
consequence of these properties is that there is a region at
intermediate radii (between 0.3 and 3~pc) where the curve can be
fitted by a power-law with an exponent of $a\sim$1.6-1.7. This is the
observational result quoted by \citet{Kauffmann+10} and
\citet{Lombardi+10}.

In order to understand the origin of this behavior, we first note
that, by construction, the maximum possible value of the slope
$\alpha$ of the $M-R$ curve is 2: at a given threshold, each cloud has
a particular size and mass, i.e., each cloud has a particular point in
the $M-R$ diagram, but all of them fall along the same straight line with
slope of 2, as a consequence of the third Larson's relation.  For
every cloud, smaller sizes and masses are obtained when we increase
the extinction threshold.  To visualize this, pay attention to any
pair of points with the same symbol in Fig. 1 by \citet{Lombardi+10}.
In this case, we are moving from one point along a line with slope of
2, to another point, with smaller mass and size, along another line
with slope of 2, but with a larger intercept.  Then, for a single
cloud, their connecting points at different thresholds are shifted
along lines with slopes smaller than two.  The limit case occurs when
the maximum column density of the cloud is reached, and it is in this
case when $\alpha\to 2$.

We now calculate the intra-cloud $M-R$ relation, for clouds with
different functional forms of their \Npdf s: lognormal, power-law and
the combination of these two.  Adopting the definition
$R\equiv\sqrt{S/\pi}$, as in most observational works on MCs, we can
calculate the mass-size relation for a single cloud from the \Npdf:
the mass will be given by eq. (\ref{eq:masa}), while the size will be
proportional to the square root of eq.  (\ref{eq:surface}).

\begin{figure}
\includegraphics[width=1.\hsize]{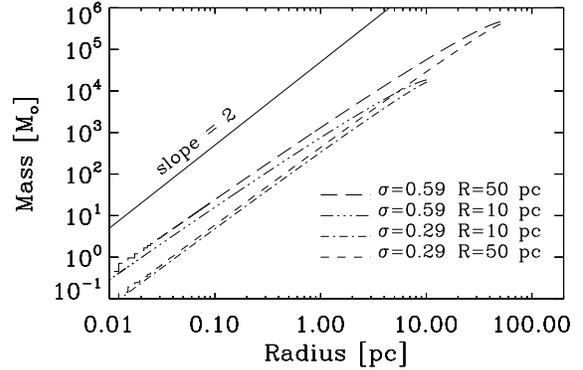}
\caption{\label{fig:mass_size_lognormal} $M-R$ diagram for four
  lognormal \Npdf s.  In all cases, $A_{\rm max}$, the peak of the
  distribution, occurs at $A_{\rm max}=0.1$.  The size of the cloud, and the
  standard deviation of the Gaussian are indicated. Lower panel: Same
  diagram for different power law \Npdf s.  }
\end{figure}

Fig.~\ref{fig:mass_size_lognormal} shows the mass-size diagram for
four lognormal functions: two with sizes similar to those of low mass
clouds ($R=10$~pc, e.g., the Pipe nebula) and two with sizes similar
to those of large-mass clouds ($R=50$~pc, e.g., Orion).  Each one of
these cases have two different values of $\sigma$: 0.29 and 0.59,
which are the extreme values of the standard deviation $\sigma$ of the
lognormal fits obtained by \citet{Kainulainen+09}.  The solid straight
line running across the diagram has a slope of 2.  As we can notice,
the slope of the mass-size relation at small radii approaches to 2, as
in Fig.~2 by \citet{Lombardi+10}.

\begin{figure}
\includegraphics[width=1.\hsize]{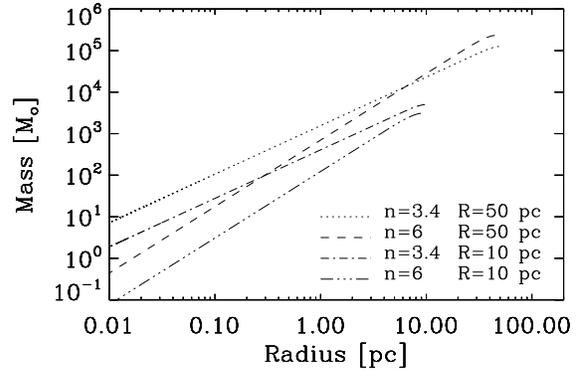}
\caption{\label{fig:mass_size_powerlaw} $M-R$ diagram for four
  power-law \Npdf s.  In all cases, $A_{\rm max}$, the peak of the
  distribution, occurs at $A_{\rm max}=0.1$. The size of the cloud,
  and the index of the power-law at large extinctions $A$ are as
  indicated.  In all cases, $q=0$. Note that the index $q$ for low
  extinctions at large radius. Thus we do not make plots for other
  values of $q$. The slope of these lines follows eq.~(\ref{eq:a_n}).}
\end{figure}

In Fig. \ref{fig:mass_size_powerlaw} we show the $M-R$ diagram for
four clouds with two power-law \Npdf s.  Again, there are two cases
with size of 10~pc, and other two with 50~pc.  The indexes in the high
extinction wing of the \Npdf\ are $n=3.5$ and 6 ($m=2.5$ and 5).  In
all cases, the index of the low extinction wing of the \Npdf\ is
$q=0$. We note that a very steep power-law \Npdf\ is needed, with an
index of $n=6$ ($m=5$), in order to reproduce the slope of $1.6$
reported by \citet{Lombardi+10} for the $M-R$ diagram at intermediate
radii.
 
\begin{figure}
\includegraphics[width=1.\hsize]{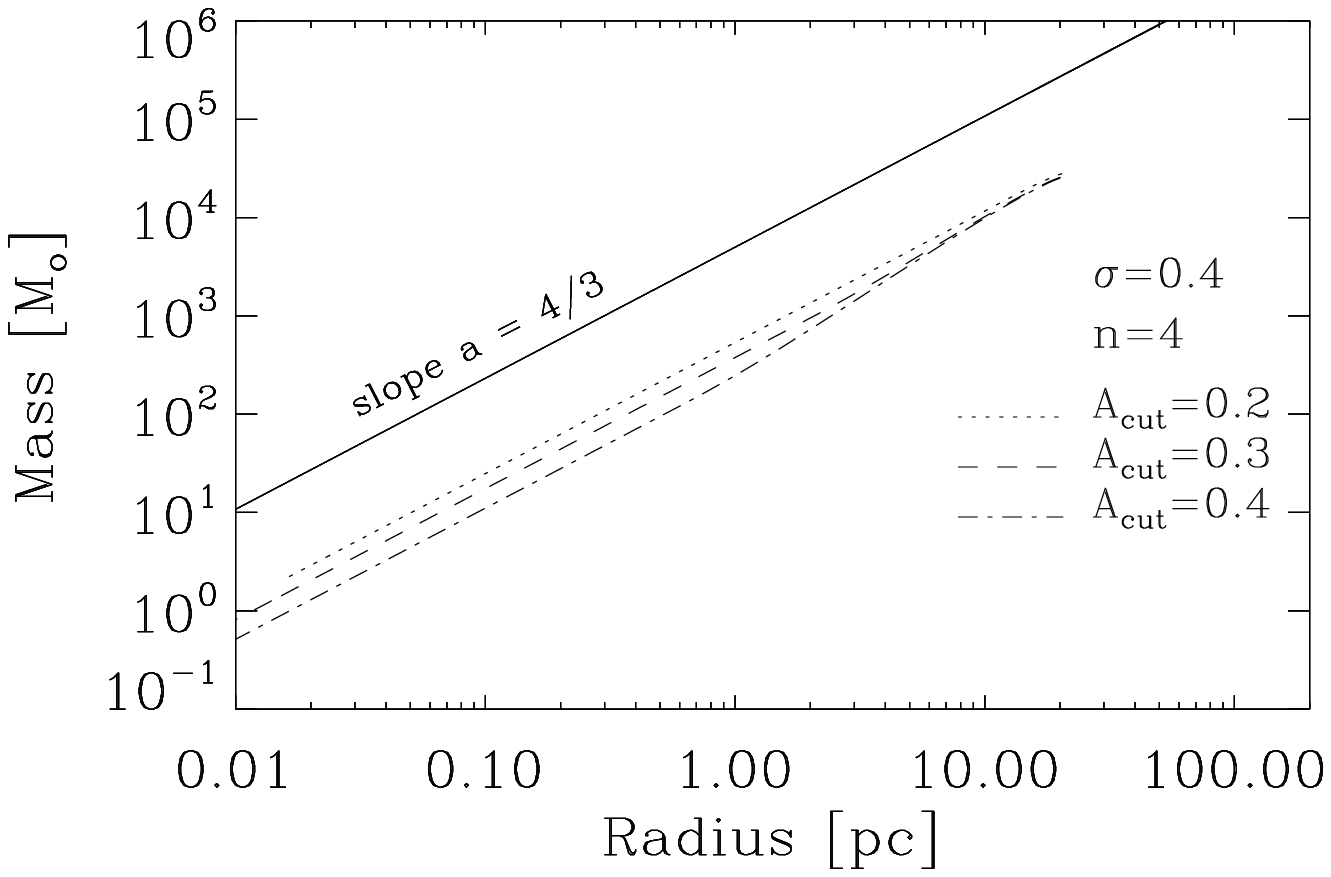}
\includegraphics[width=1.\hsize]{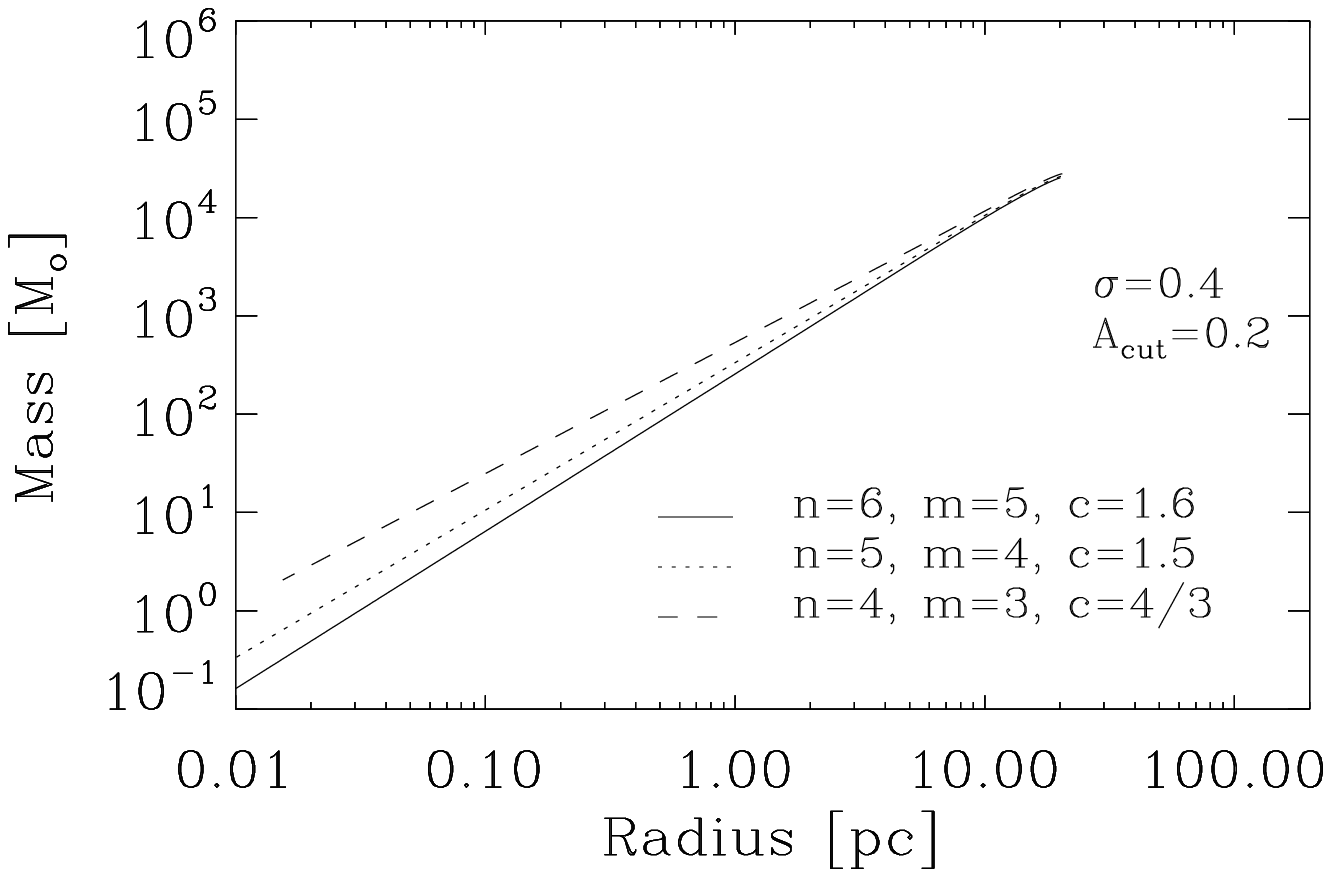}
\includegraphics[width=1.\hsize]{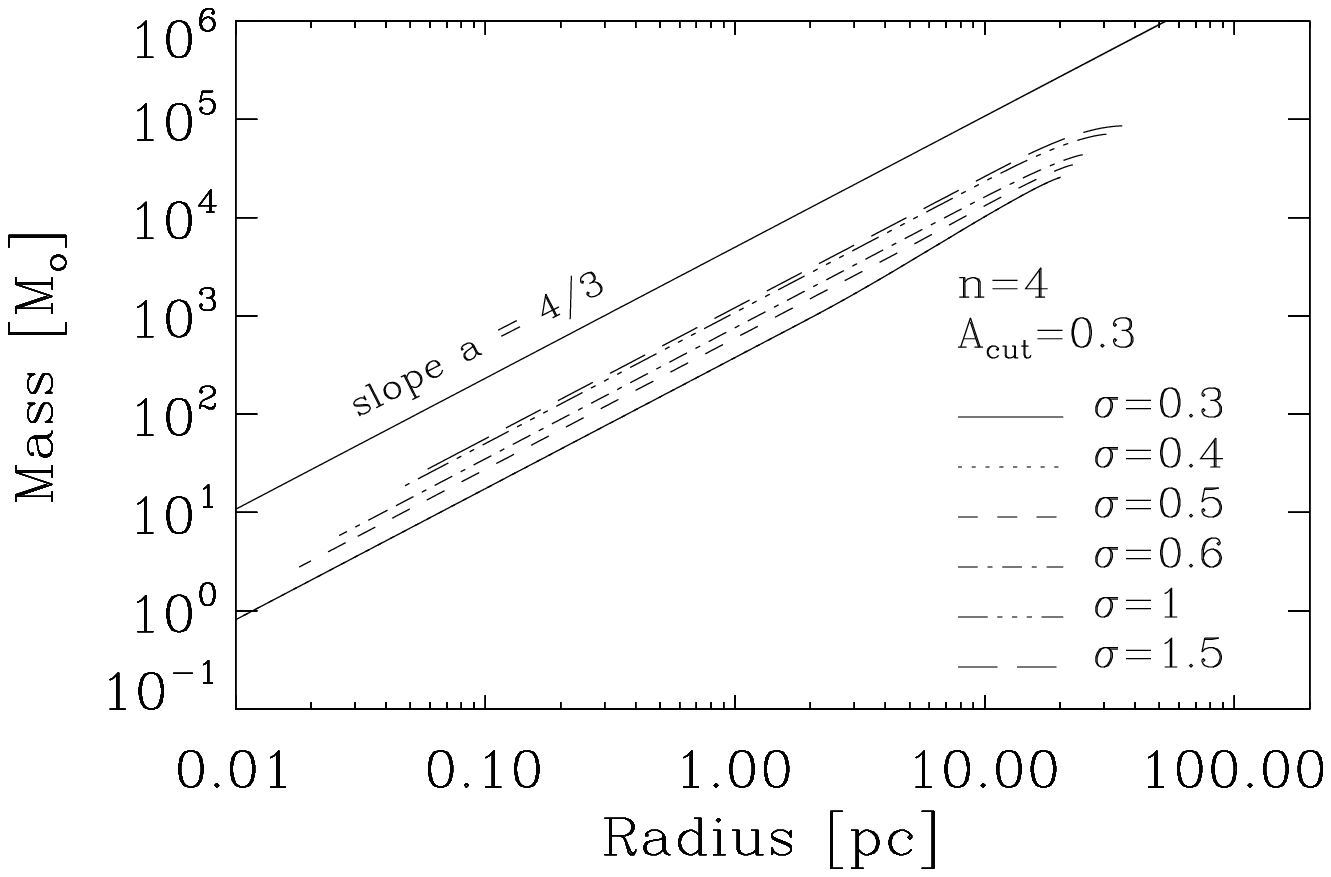}
\caption{\label{fig:mass_size_lognormal+powerlaw} $M-R$ diagram for
  lognormal$+$power-law \Npdf s. In the upper panel, we vary the
  extinction value where the transition between the lognormal and
  power-law regime occurs, $A_{\rm cut}$. In the middle panel we vary
  the slope of the power-law regime, $n$. Finally, in the lower panel,
  we vary the width of the lognormal function, $\sigma$.}
\end{figure}

In Fig.~\ref{fig:mass_size_lognormal+powerlaw}, the mass-size diagram
for clouds with a lognormal function at low column densities, and a
power-law regime at large extinctions \Npdf. The fiducial parameters
are: $A_{\rm cut}=0.2$, $n=4$, and $\sigma=0.4$.  In the upper panel
we vary the value where the transition between the lognormal and the
power-law occurs: $A_{\rm cut}=0.2$, 0.3 and 0.4.  In the middle
panel we vary the slope of the power-law regime: $n=6$, 5 and
4. Finally, in the lower panel we vary the width of the lognormal:
$\sigma=$0.3, 0.4, 0.5, 0.6, 1 and 1.5.

What these figures show us is that the width of the lognormal function
shifts the relation: the wider the \Npdf, the larger the mass at a
given radius.  Similarly, the slope $n$ of the power-law in the
\Npdf\ determines the slope $a$ of the $M-R$ relation: Assuming a
power-law \Npdf, from eqs.  (\ref{eq:mass_singlepowerlaw}) and
(\ref{eq:surface_singlepowerlaw}), we can solve for $R=\sqrt{S/\pi}$,
and show that the index $a$ in eq.  (\ref{eq:MRsinglecloud}) will be
related to the index $n$ of the high extinction wing of the \Npdf\ by

\begin{equation}
  a = {4-2n\over 1-n}.
\label{eq:a_n}
\end{equation}

\begin{figure}
\includegraphics[width=1.\hsize]{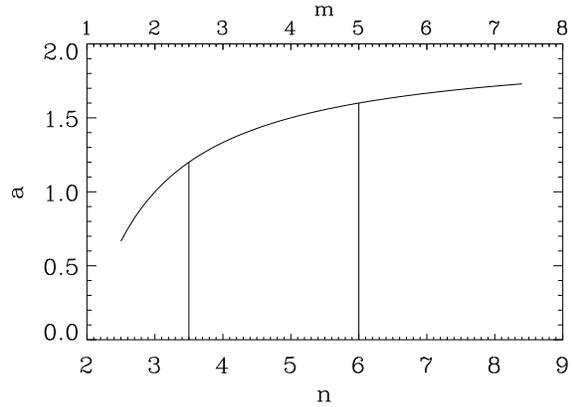}
\caption{\label{fig:a_n} $n-a$ diagram, according to
  eq.~(\ref{eq:a_n}), where $a$ stands for the exponent of the $M-R$
  relation, while $n$ stands for the exponent of the power-law of the
  \Npdf\ at large extinctions. We recall that the slope measured from
  the \Npdf\ when plotted in log-log scale is given by $m=n-1$, as
  given in the upper $x$-axis. Notice that value $a=2$ is forbidden,
  according to eq.~\ref{eq:a_n}.}
\end{figure}

Fig.~\ref{fig:a_n} shows the $n-a$ diagram, as given by
eq.~(\ref{eq:a_n}). We notice that $a$ moves softly from $\sim 1.2$ to
1.6 for realistic values of the power-law ($n=3.5-6$, or
$m=2.5-5$). Note furthermore that value $a = 2$ is forbidden,
according to eq.~(\ref{eq:a_n}).

All together, our results and the observational results present the
following puzzle: while most of the observed clouds exhibit a
power-law distribution \Npdf\ for large extinctions, why such clouds
exhibit an $M\propto R^2$ relation at small radii, if the value $a=2$
is forbidden for power-laws at large extinctions?  In principle, this
value will be allowed only for the few clouds that exhibit lognormal
\Npdf s, as Coalsack, California, the Pipe, Lupus V, LDN 1719.

In order to explain this apparent contradiction, lets assume that we
have a \Npdf\ with a shape that changes continuously, such that it can
be approximated as a sequence of power-laws in small intervals of $A$:

\begin{equation}
p(A) dA = p_i \biggl( {A\over A_i} \biggr)^{-n_i} dA \quad \mbox{for
    $A \epsilon (A_i,A_{i+1})$, $A_i<A_{i+1}$ }
\label{eq:Npdf_sequencial}
\end{equation}

We now assume that the \Npdf\ decreases rapidly, such that the mass at
a given threshold is dominated by the low-density region (this
assumption is fulfilled by the observed clouds, since this is one of
the conditions needed in order to obtain the third Larson relation, as
we showed in the previous sections).  In this case, we can integrate
out eqs.~(\ref{eq:surface}) and (\ref{eq:masa}) only for the small
interval in which the power-law applies, i.e.,

\begin{eqnarray}
  S(A_0) & = & S_{\rm tot}\ \int_{A_0}^\infty \ p(A)\ dA \simeq
  \nonumber\\ & & S_{\rm tot}\ \int_{A_0=A_i}^{A_{i+1}} \ p_i
  (A/A_i)^{-n_i} dA 
\label{eq:surface_sequence}
\end{eqnarray}
and

\begin{eqnarray}
  M(A_0) & = & S_{\rm tot} \mu m_H\ \beta \int_{A_0}^\infty
  A\ p(A)\ dA \simeq \nonumber \\ & & S_{\rm tot} \mu m_H\ \beta
  \int_{A_0=A_i}^{A_{i+1}} A\ \ p_i (A/A_i)^{-n_i} dA .
\label{eq:masa_sequence}
\end{eqnarray}
Solving these equations we obtain

\begin{equation}
  S(A_0)  =  {S_{\rm tot}\ A_i\ p_i\over 1-n_i}
  \biggl[\biggl({A_{i+1}\over A_i}\biggr)^{1-n_i} - 1 \biggr]
\label{eq:surface_sequence_solution}
\end{equation}
and

\begin{equation}
  M(A_0) = {S_{\rm tot}\ \mu m_H\ \beta\ p_i\ A_i^2\over 2-n_i}
  \biggl[\biggl({A_{i+1}\over A_i}\biggr)^{2-n_i} - 1 \biggr]
\label{eq:masa_sequence_solution}
\end{equation}
Since we have defined $S=\pi R^2$, we can write the ratio $A_{i+1}/A_i$ in
terms of $R$ from eq.~(\ref{eq:surface_sequence_solution}):

\begin{equation}
  {A_{i+1}\over A_i}= \biggl\{ 1+{1-n_i\over S_{\rm tot}\ A_i\ p_i} \pi R^2
  \biggr\}^{1/(1-n_i)}
\label{eq:ratioA_0:A_i}
\end{equation}
and insert this value in eq.~(\ref{eq:masa_sequence_solution}) to
obtain

\begin{equation}
  M(A_0)  =  {S_{\rm tot}\ \mu m_H\ \beta p_i A_i^2\over 2-n_i}
  \biggl\{\biggl({1-n_i\over S_{\rm tot} A_i p_i} \pi
  R^2 +1 \biggr)^\delta -1 \biggr\}
\label{eq:masa_vs_R}
\end{equation}
with 

\begin{equation}
  \delta={2-n_i\over 1-n_i} ,
\label{eq:delta}
\end{equation}
where $\delta < 1$.

Lets now analyze the limit cases. For large radii (low extinctions),
$\pi R^2\sim S_{\rm tot}$.  Note that the ratio $(1-n_i)/A_ip_i \ge
25$: observations by \citet{Kainulainen+09} show that the typical
values of the parameters are: $3.5\le n_i\le 6$, $0.1\le A_i\le
2$,(recall that $A_V\sim 10 A_k$, and our $A_i$ are assumed to be in
the $K$ band) and $p_i\sim 1$ for $A_i\sim 0.1$, and $p_i\leq 10^{-3}$
for $A_i\sim 2$ \citep[see \Npdf s in][]{Kainulainen+09}.  Thus, the
term in parenthesis becomes much larger than unity, and then,

\begin{equation}
  M(A_0) \propto R^{2\delta} 
\label{eq:masa_vs_R_largeR}
\end{equation}
for large radii, recovering the solution for the single power-law
case, eq.~(\ref{eq:a_n}).  Note that $2\delta < 2$, since $\delta <
1$.  As an example, lets take for instance the \Npdf\ of Lupus I
cloud, shown in Fig.~4 by \citet{Kainulainen+09}.  The slope for
$A_V\leq$~1 is approximately $m=3.3$. Thus, $n=4.3$ and
$2\delta=1.13$. This is the value of the slope that can be measured
from Fig.~1 in \citet{Lombardi+10} for the black circles that denote
Lupus 1.

Lets now analyze the opposite case. For small radii, 

\begin{equation}
{1-n_i\over S_{\rm tot}\ A_i\ p_i} \pi R^2 << 1
\label{eq:small_radii_condition}
\end{equation}
and thus the expansion of eq.~(\ref{eq:masa_vs_R}) in Taylor series
gives the relation

\begin{equation}
  M(A_0) = \mu m_H\ \beta\ A_i\ \pi R^2 
\label{eq:masa_vs_R_smallR}
\end{equation}
i.e., for small radii all the slope of the $M-R$ curves should
approach to two, the limit case.  For intermediate radii, however, the
solution must be given by eqs.~(\ref{eq:masa_vs_R}) and
(\ref{eq:delta}).  

In summary, as we can see from
eqs.~(\ref{eq:masa_vs_R})--(\ref{eq:masa_vs_R_smallR}), the intracloud
mass-size relation is completely defined by the shape of the \Npdf,
and transitions from a slope of 2, at small radii, towards flatter
slopes, at large radii.

\section{Discussion}\label{sec:discussion}

\subsection{The inter-cloud relation}

Early discussions of the importance of the apparent constancy of the
column density of MCs found by \citet{Larson81} emphasized the limited
column densities probed by then-available $^{12}$CO observations
\citep{Kegel89}.  Since that time, other tracers of column density
such as near-infrared extinction with larger dynamic range, have been
used to derive similar results.  In particular, \citet{Lombardi+10}
and \citet{Beaumont+12} have found $M \propto$~area at differing
thresholds, though with differing normalization.

In the first sections of this work we showed how a wide variety of
\Npdf s can reproduce the inter-cloud Larson's 3rd law, i.e., can
produce $M \propto$~area relations for multiple clouds.  Furthermore,
we showed that if one could in principle threshold the volume density,
one would get a $M \propto$~volume relation.  This strongly suggests
that these types of relations, rather than universal laws, are mostly
an artifact of the thresholding process. 

Operationally, cloud masses are obtained by integrating or summing
column densities over the cloud area.  It is thus not surprising that
cloud masses are observed to correlate with areas, specially, if most
of the mass is at low densities, i.e., occupying most of the area.  If
then the \Npdf\ is rapidly decreasing with increasing column density,
this means that most of the column density and thus most of the mass
is near the threshold.  Thus one automatically gets a cloud mass
linearly proportional to its area.  Similarly, if one were able to
measure densities directly, the same principle results in mass
linearly proportional to volume.

The significance of the mass-area relationship then reduces to
understanding why the mean or peak of the \Npdf\ seems not to vary
much from region to region \citep{Beaumont+12}, and why the \Npdf\ has
a rapid decrease with increasing column density in the thresholds of
interest.  We argue that there are good physical reasons for this.
The first is that, at least in the Solar Neighborhood, an extinction
of $A_V\sim 1-2$, equivalent to a hydrogen column density of
$N\sim$\diezala{21} cm\alamenos 2, is approximately what is needed to
shield CO from the dissociating photons of the interstellar radiation
field \citep{vanDishoeckBlack88, vanDishoeckBlacke98}.  It is
difficult to set thresholds for molecular clouds below this average
value \citep[which corresponds to $A \sim 0.1$ at K-band, the lowest
  level considered by][]{Lombardi+10}, since as it would not be a CO
cloud and these studies are focused towards MCs.

Secondly, we argue that most of the mass of clouds is at low density,
because the shielding value is roughly the column density at which one
expects self-gravity to become important \citep{FrancoCox86, HBB01}.
Numerical simulations \citep[e.g.,][]{BP+11b} indicate that the slope
of the \Npdf\ at large column densities $N$ is steeper than $3$, as a
result of gravitational collapse.  More specifically, in
non-spherical, non-uniform geometries gravitational collapse can be
highly non-linear as a function of position \citep{BH04, HBH12}, such
that the dense material is a small fraction of the total cloud mass.

\citet{Beaumont+12} concluded that the ``universality'' of Larson's
third relation is a result of having the dispersion in the mean value
of the column density be smaller than the dispersion in cloud areas.
Our argument is that this occurs because the large column density
regions of clouds have a small filling factor as a consequence of the
gravitational contraction and collapse, producing the mass
$\propto$~area result.  We agree with \citet{Beaumont+12} that
therefore the \Npdf\ itself is much more instructive about cloud
structure than the mass-area ``law'', and that the power-law or other
high-column density ``tails'' on \Npdf s is a result of gravitational
collapse \citep{BP+11b}.  This is consistent with our previous
arguments that modern versions of Larson's second relation
\citep{Heyer+09} imply global (and local) gravitational collapse
\citep{BP+11a}, not Virial equilibrium.

\subsection{The intra-cloud relation}

{Although the original mass-size relation by \citet{Larson81} $M\propto
R^2$ was obtained for a set of clouds, i.e., in terms of the
inter-cloud relation, different authors had noticed that this relation
does not holds for a single cloud.  The typical values of the exponent
used to be smaller, typically $M\propto r^{1.6-1.7}$
\citep{Kauffmann+10, Lombardi+10}, although it should be recognized
that the diagrams presented by \citet{Lombardi+10} show that the
relation is a curved line, rather than a power-law.  The results
presented in the present paper shows that the \Npdf\ of the cloud
determines the slope of the $M-R$ relation, and it goes from 2 as a
limit case for small radii, and flattens for larger radii.
}

\section{Conclusions}\label{sec:conclusions}

By using simple functional forms of the column density probability
distribution function (\Npdf), we have explored the significance of
the third Larson scaling relation for molecular clouds.  We show that
for a set of clouds, this relation is an artifact of thresholding the
surface density, as long as the form of \Npdf\ peaks near or below the
threshold used, and that the mass is a steeply decreasing function of
increasing column density.  We argue that the physical reasons for
these features of the column density distribution are that there is a
minimum column density to shield CO, and that this minimum value is
not far from that at which the molecular gas becomes Jeans unstable,
resulting in cloud collapse to form stars.

{We also showed that for a single cloud, the slope of the mass-size
relation approaches to 2 at small radii, and flattens for larger
radii, in agreement with observations.  The detailed shape of this
curve depends on the detailed shape of the \Npdf.
}

\section{Acknowledgments}

This work was supported by UNAM-PAPIIT grant number IN103012 to JBP
and IN102912 to PD, and by US National Science Foundation grant
AST-0807305 to LH.  We have made extensive use of the NASA-ADS
database.


\label{lastpage}

 \end{document}